\documentclass[conference]{IEEEtran}
\IEEEoverridecommandlockouts
\usepackage{color}
\usepackage{graphicx}  % Written by David Carlisle and Sebastian Rahtz
\usepackage{psfrag}    % Written by Craig Barratt, Michael C. Grant,
\usepackage{algorithm}
\usepackage{algorithmic}
\usepackage{amsfonts,amssymb}
\usepackage{amsthm}
\usepackage{amsmath}
\usepackage{amstext}
\usepackage{bm}
\usepackage{subfigure}
\usepackage{hyperref}
\usepackage{CJK}
\usepackage{url}
\usepackage[square, comma, sort&compress, numbers]{natbib}
\usepackage{epstopdf}

\begin{document}

\title{Learning-Based Robust Resource allocation for D2D Underlaying Cellular Network}

\author{Weihua Wu, Runzi Liu, Qinghai Yang,  and Tony Q. S. Quek, \emph{Fellow, IEEE}
\thanks{*This work was supported in part by the NSF China under Grant 61801365, 61701365 and 61971327, in part by the China Postdoctoral Science Foundation under Grant 2018M643581, in part by the National Natural Science Foundation of Shaanxi Province under Grant 2019JQ-152, in part by Postdoctoral Foundation in Shaanxi Province of China, and the Fundamental Research Funds for the Central Universities.}
\thanks{W. Wu and Q. Yang are with State Key Laboratory of ISN, School of Telecommunications Engineering, Xidian University, No.2 South Taibai Road, Xi'an, 710071, Shaanxi, China. (Email:whwu@xidian.edu.cn).}
\thanks{R. Liu is with School of Information and Control Engineering, Xi'an University of Architecture and Technology, Xi'an 710055, China.}
\thanks{T. Q. S. Quek is with the Information Systems Technology and Design Pillar, Singapore University of Technology and Design, Singapore 487372 (e-mail: tonyquek@sutd.edu.sg).}

}
\maketitle

%while guaranteeing the minimum signal-to-noise ratio (SINR) requirement of certain throughput for V2I communications and the probabilistic quality-of-service (QoS) constraints for V2V communications

\begin{abstract}

In this paper, we study the resource allocation in D2D underlaying cellular network with uncertain channel state information (CSI). For satisfying the diversity requirements of different users, i.e. the minimum rate requirement for cellular user and the reliability requirement for D2D user, we attempt to maximize the cellular user's throughput whilst ensuring a chance constraint for D2D user. Then, a robust resource allocation framework is proposed for solving the highly intractable chance constraint about D2D reliability requirement, where the CSI uncertainties are represented as a deterministic set and the reliability requirement is enforced to hold for any uncertain CSI within it. Then, a symmetrical-geometry-based learning approach is developed to model the uncertain CSI into polytope, ellipsoidal and box. After that, we derive the robust counterpart of the chance constraint under these uncertainty sets as the computation convenient convex sets. To overcome the conservatism of the symmetrical-geometry-based uncertainty sets, we develop a support vector clustering (SVC)-based approach to model uncertain CSI as a compact convex uncertainty set. Based on that, the chance constraint of D2D is converted into a linear convex set. Then, we develop a bisection search-based power allocation algorithm for solving the resource allocation in D2D underlaying cellular network with different robust counterparts. Finally, we conduct the simulation to compare the proposed robust optimization approaches with the non-robust one.

\vspace{5pt}
\textbf{\emph{Key Terms}}: D2D communications, resource allocation, robust optimization, chance constraint, SVC.

\end{abstract}


%for peerreview papers, inserts a page break and creates the second title.
% Will be ignored for other modes.

\section{Introduction}

\newtheorem {theorem}{\textbf{Theorem}}
\newtheorem {lemma}{\textbf{Lemma}}
\newtheorem {remark}{\textbf{Remark}}
\newtheorem {definition}{\textbf{Definition}}

The throughput requirements in cellular networks have been exponentially increasing in these years \cite{D.Chmieliauskas}. Device-to-device (D2D) is considered as a promising technique to satisfy this demand so that it received much attentions over the last few years \cite{K.Doppler}. With the deployment of D2D, the cellular users can communicate with each other directly without bypassing the base station (BS). This paradigm entails a wide range of benefits, such as reusing gain, proximity gain and hop gain \cite{D.FengCM,H.SunICC}. Due to these advantages, D2D technique is widely used in the fields of Internet of vehicles, Internet of things, industrial Internet and so on.

There exist two paradigms for the D2D coexisting with cellular users:  underlaid paradigm and dedicated paradigm \cite{H.SunICC}. In the underlaid paradigm, the same resource used by a traditional cellular user can be simultaneously used by a D2D pair when it is located in a sufficiently distant part of the cell. Hence, it can provide services for more users without adding the bandwidth. However, there exists cross-tier interference between D2D users and cellular users, which seriously degrades the capacity of each transmission link. In the dedicated paradigm, the BS should allocate orthogonal resources to D2D users and cellular users. Thus, the cross-tier interference can be effectively cancelled and the link performance is enhanced. However, due to the limited resources, the number of users that the network can accommodate is strictly limited. Therefore, in order to serve more users and obtain higher spectrum efficiency, it is of great significance to implement the underlaid coexistence between D2D users and cellular users.

In order to fully exploit the potential benefits of underlaid D2D communications, it is necessary to provide a judicious assignment of the spectrum resource (e.g. time slot or resource block) to D2D users and design a prudent power control mechanism that avoids the detrimental interference to cellular users. This is the case in \cite{W.ZhaoCL,W.LaiTVT}, where the resource allocation approaches were developed to work out the solution of subchannel assignment and power distribution.  The unified resource management schemes in \cite{D.ZhaiTWC,Z.SUNWCSP} achieved the coexistence of D2D and cellular by jointly optimizing mode selection, resource allocation, and power control. All of the previously mentioned works assume the perfect channel state information (CSI) is available at the BS. From the practical prospective, collecting the D2D's CSI requires a lot of cooperation between D2D pairs and the cellular users, then substantial amount of communication and latency overhead will be added before the BS receives the value. As a result, the BS can only obtain the uncertain value about the CSI of D2D. The uncertainties of CSI render the deterministic optimizations in \cite{W.ZhaoCL,W.LaiTVT,D.ZhaiTWC,Z.SUNWCSP} unreliable. It has been widely demonstrated in \cite{L.WangTCOM}\cite{S.Parsaeefard} that even a slight perturbation on the CSI can greatly influence the system performance, such as leading to suboptimal objective or the violation of quality-of-service (QoS) constraint.

Motivated by urgent requirement of handling the uncertain value about the CSI of D2D communications, stochastic optimization and robust optimization methods have received many attentions in recent years. With the deployment of stochastic optimization, the works in \cite{L.LiangTCOM,X.LiIOT,C.GuoJSAC,W.SunTWC} tackled the uncertainties of CSI by optimizing the expected link capacity. Besides that, \cite{Mohamed} transformed the QoS requirement with uncertain CSI into a closed-form expression under the given statistical distribution of the uncertainty. Evidently, the stochastic optimization technique entails complete distribution information about the uncertain CSI. This may be unrealistic in practical wireless networks. For the robust optimization, there exist two approaches: the Bernstein approximation approach \cite{Mathematical,N.MokariTMC} and the worst-case optimization approach \cite{L.WangTCOM,Y.HaoTCOM}. The former assumes that the uncertain channels are uncorrelated and that all of the CSI values are independent with each other. Then, it optimizes the network utility with guaranteeing the QoS constraint with certain probability. However, we should note that the independent distribution assumption in Bernstein approximation is too idealistic and that it loses a lot of useful correlation information between different channels, so that it may lead to over conservative solution. On the other hand, the worst-case optimization takes a deterministic and set-based approach to grantee the worst-case performance and satisfy the constraint under any realization of CSI in a closed uncertainty region. For the case of unbounded uncertainty, the worst case approach is still applicable by satisfying the QoS constraint with a certain probability \cite{G.ZhengTWC}. Due to these appealing characteristics, this paper exploits the worst case approach to address the resource allocation in D2D underlaying cellular network.

As for the worst case robust optimization in D2D networks, there already exist some pioneering works \cite{Y.HaoTCOM,H.XuTVT,Y.MaACCESS}. More specifically, \cite{Y.HaoTCOM,H.XuTVT} focused on maximizing the worst-case system utility of both cellular and D2D links under CSI impairment. The work in \cite{Y.MaACCESS} utilized the worst-case approach to find the tractable forms about the chance constraint and then derived the semi-closed expression for the power allocation in D2D networks. However, all of these works deploy the robust optimization based on the fixed-size uncertainty CSI set. Actually, this assumption is excessively idealistic. In practical wireless network, the size and shape parameters of the uncertainty set would vary with the network conditions. Thus, constructing the uncertainty region to include probable realizations of uncertain CSI is a paramount ingredient in robust optimization.

In this paper, we consider the resource allocation in D2D underlaying cellular network. In order to improve the spectrum efficiency, the D2D users reuse the uplink spectrum resource of cellular users. We assume that the CSI connect to BS can be perfectly obtained since it can be directly estimated by the BS, while the CSI of D2D links are reported to BS with errors. Because D2D communication is usually used to support the emergency services, such as the emergency electronic brake lights in vehicle networks \cite{G.NaiKACCESS}, we define a chance constraint for the D2D QoS requirement. On the contrary, the minimum QoS requirement constraint is defined for the cellular user since it is often used for the non-safety related service in most of case. Different from the previous robust optimization approaches, where the uncertainties are assumed to being independently and symmetrically distributed, we consider a more realistic scenario that the uncertainties are intertwined and asymmetric. Then, we investigate the different robust optimization approaches  for solving the resource allocation problem in D2D underlaying cellular network. The main contributions of this paper can be summarized as follows:

\begin{itemize}
  \item A robust resource allocation framework is developed for solving the highly intractable chance constraint about D2D QoS requirement. Specifically, the CSI uncertainties are modeled as a deterministic (uncertainty) set, and the chance constraint is satisfied by enforcing the QoS requirement to hold for any uncertain CSI within it
  \item We develop a symmetrical-geometry-based learning approach to construct the uncertainty set from the samples of uncertain CSI. More specifically, the uncertainties are constructed as polytope, ellipsoidal and box, respectively. Then, the robust counterparts of the chance constraint corresponding to the different uncertainty sets are derived. It should be noted that all of the robust counterparts are definitely computationally convenient for the convex optimization tools.
  \item To overcome the conservatism of the symmetrical-geometry-based uncertainty sets, we develop a support vector clustering (SVC)-based approach to construct the uncertainties as a compact convex set. We should note the SVC-based uncertainty set is asymmetric and can tightly enclose the uncertainties without any superfluous space. Moreover, the induced robust counterpart is the tractable linear convex set. In this case, we also found that the robust optimization works as a foundation for bridging modern machine learning tools into the resource allocation in wireless network.
  \item A bisection search-based power allocation algorithm is developed for solving the power allocation problem under diversity robust counterparts with low polynomial-time complexity.
\end{itemize}

The rest of this paper is organized as follows. In Section II, we give the system model and problem formulation. The basic uncertainty set induced robust resource allocation is illustrated in Section III. Section IV investigates the kernel learning induced robust allocation.  Afterwards, the simulation results are illustrated in Section V. Finally, Section VI concludes the paper.

\section{System Model}

This section provides the network model and channel uncertainty model of the  D2D underlaying cellular network, followed by the problem formulation about the power allocation under diversity QoS requirements.

\begin{figure}
\begin{center}
\includegraphics[width=3.3in,height=1.7in]{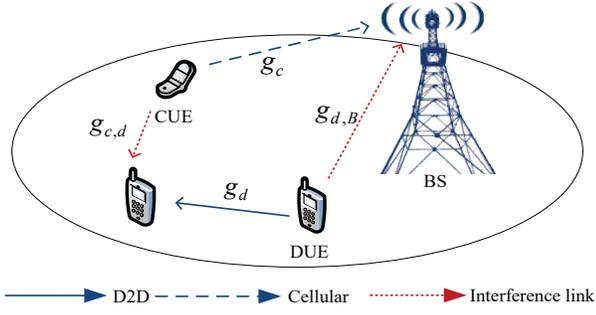}
\label{fig2}\caption{System model.}
\end{center}
\end{figure}

\subsection{Network Model}

We consider a D2D underlaying cellular network as shown in Fig. 1. In the considered system model, the users, denoted as CUE, communicate with the BS through the cellular link. Because the uplink resources of cellular links are less intensively used and the BS has a stronger ability to manage the interference than users, the uplink spectrum resource of CUE is reused by the D2D users, denoted as DUE, for improving the spectrum utilization efficiency. In order to reduce the complexity brought by the complicated inter- and intra-cell interference, we assume that one DUE is only allowed to access the spectrum of one CUE and that the spectrum of one CUE can only be reused by one DUE. It has been widely proved in \cite{L.LiangTCOM,X.LiIOT,C.GuoJSAC} that the optimal spectrum reusing pattern can be easily obtained by using the Hungarian method \cite{Hungarian} based on the optimal power allocation of all possible reusing pairs between CUEs and D2Ds. Thus, this paper does not study the spectrum allocation anymore, but focuses on the challenging power allocation for a possible reusing pair. Let $p_{c}$  denote the transmit power of CUE and $p_{d}$ denote the transmit power of the transmitter in D2D pair. Then, the channel power gain of D2D link is assumed to follow
\begin{eqnarray}
g_{d}=h_{d}\beta_{d} \varpi D^{-\iota}\triangleq |h_{d}|^{2} \alpha_{d},
\end{eqnarray}
where $\alpha_{d}=\beta_{d} \varpi D^{-\iota}$ represents the large-scale slow fading channel gain,  $h_{d}$ is the small-scale fast fading power component, $\varpi$ is the pathloss constant, $D$ is the distance of the DUEs in pair, $\iota$ is the decay exponent, and $\beta_{d}$ is log-normal shadow fading random variable with a standard deviation $\xi$. Then, the channel power gain between CUE and BS $g_{c}$, the corsstalk channel power gain between CUE and DUE $g_{c,d}$, and the crosstalk channel power gain between DUE and BS $g_{d,B}$  have the similarly definitions.

We assume that the CSI between BS and users i.e., $g_{c}$ and $g_{d,B}$ can be perfectly obtained because they can be directly estimated by the BS. For the CSI between different users, i.e., $g_{d}$ and $g_{c,d}$, the large-scale fading is accurately known since they usually dependent on the locations which vary on a slow scale. On the contrary, due to the Doppler effect \cite{X.LiIOT}, partial CSI acquisition \cite{Y.ShiTSP} and the delays in CSI feedback \cite{M.A.TIT}, there inevitably exists uncertainty in the small-scale CSI. In this work, the additive error \cite{K.YangIOT,L.LiangWCL} model of the channel imperfection is adopted, i.e.,
\begin{eqnarray}
|h_{d}|^{2}=\delta^{2} |\hat{h}_{d}|^{2}\!+\!(1-\delta^{2})|e_{d}|^{2},
\end{eqnarray}
where $\hat{h}$ is the estimated channel gain, $e$ is the estimation error and $\delta$ $(0<\delta<1)$ is the channel estimation error coefficient. It can be easily understood that $\delta=0$ means that no CSI is acquired at all, while $\delta=1$ indicates the perfect channel estimation. Most of the current works consider that both $\delta$ and $e$ are known to the transmitter and that the estimation errors of different channels are independent with each other. This paper considers a more practical and general scenario where all of the distribution information about them are unknown and the estimation errors of different channels are correlated and intertwined with each other. Moreover, the underlying distribution of channel errors may be intrinsically complex and variable with the evolution of wireless network.

Based on the these assumptions, the received Signal to Interference plus Noise Ratio (SINR) of D2D pair can be given as
\begin{eqnarray}
\gamma_{d}\!=\!\frac{p_{d} \alpha_{d}\left(\delta^{2} |\hat{h}_{d}|^{2}\!+\!(1-\delta^{2})|e_{d}|^{2}\right)}{\sigma^{2}+p_{c}\alpha_{c,d}\left(\delta^{2} |\hat{h}_{c,d}|^{2}\!+\!(1\!-\!\delta^{2})|e_{c,d}|^{2}\right)},
\end{eqnarray}
where $\sigma^{2}$ is the power of the additive white Gaussian noise. Similarly, the received SINR at BS from the CUE can be computed as
\begin{eqnarray}
\gamma_{c}=\frac{p_{c}\alpha_{c}h_{c}}{\sigma^{2}+p_{d}h_{d,B}\alpha_{d,B}}.
\end{eqnarray}

\subsection{Problem Formulation}

In D2D underlaying cellular network, we assume that the CUE and DUE have different QoS requirements. Because the D2D communication is usually used for safety related service \cite{W.SunTWC}, it has the objective of reducing the outage probability. On the contrary, the main objective of CUE is to enable a more efficient and comfortable entertainment experience, the minimum QoS requirement is necessary for them. Based on these considerations, the problem can be formulated as
\begin{subequations}
\begin{eqnarray}
\max\limits_{ p_{c},p_{d} }&& B\log_{2}(1+\gamma_{c}) \\
\textrm{s.t.}&&   \gamma_{c} \geq \gamma_{min}^{c},    \\
&&  \Pr\left\{\gamma_{d}\geq \gamma_{min}^{d}\right\}\geq 1-\epsilon,   \\
&& 0\leq p_{c}\leq P_{max}^{c}, \\
&& 0\leq  p_{d}\leq P_{max}^{d},
\end{eqnarray}
\end{subequations}
where $B$ denotes the spectrum bandwidth, $\gamma_{min}^{c}$ and $\gamma_{min}^{d}$ are the minimum required SINR for CUE and DUE, respectively, $P_{max}^{c}$ and $P_{max}^{d}$ indicate the maximum transmit power of CUE and DUE, respectively. In (5c),  $\Pr\{\cdot\}$ denotes the probability of the input, $\epsilon$ is the maximum tolerable outage probability for D2D communication. Obviously, the chance constraint induced by uncertain CSI poses a great challenge on solving the resource allocation problem. In the following, we focus on solving the power allocation problem based on the uncertain CSI.

\section{Basic Uncertainty Set Induced Robust Resource Allocation}

To avoid the complexity of directly solving the resource allocation in (5), this section employs the robust optimization approach to represent the uncertain CSI of D2D links as a high probability region (HPR) and then enforces the QoS constraint to hold for any CSI value in it. To obtain the HPR, we need to collect multiple samples of the uncertain CSI of D2D links and then learn the uncertainty set. For satisfying the chance constraint of D2D, the uncertainty set needs to cover the samples with a certain probability. If the obtained power allocation solution is feasible under all of the CSI values in the uncertainty set, the chance constraint must be satisfied. Motivated from this consideration, we first formulate problem (5) as the following approximation form
\begin{subequations}
\begin{eqnarray}
\max\limits_{ p_{c},p_{d} }&& B\log_{2}(1+\gamma_{c}) \\
\textrm{s.t.}&&   \gamma_{c} \geq \gamma_{min}^{c},    \\
&&  \mathbf{p}^{T}\mathbf{g}\geq \gamma_{min}^{d}, \mathbf{g}\in \mathcal{G},  \\
&&\textrm{(5d)},\textrm{(5e)}, \nonumber
\end{eqnarray}
\end{subequations}
where $\mathbf{p}=[p_{d}/\sigma^{2},-p_{c}\gamma_{min}^{d}/\sigma^{2}]^{T}$, $\mathbf{g}=[g_{d},g_{c,d}]^{T}\in \mathbb{R}^{2}$ and $\mathcal{G}$ is the HPR that need to be leaned. It is not difficult to understand that if $\mathcal{G}$ covers the samples of uncertain CSI $\mathbf{g}$ with $1-\epsilon$ confidence level, any feasible power allocation solution of (6) must satisfy
\begin{eqnarray}
\Pr\left\{\gamma_{d}\geq \gamma_{min}^{d}\right\}\geq \Pr\left\{\mathbf{g}\in\mathcal{G} \right\}\geq 1-\epsilon,
\end{eqnarray}
which indicates that the chance constraint is  feasible for problem (5). Based on the above discussions, learning the uncertainty set is the key component for solving the power allocation. Because $\mathbf{g}\in \mathcal{G}$ can be considered a constraint condition of problem (6), the shape of uncertainty set must be considered from the tractability in the optimization problem. Moreover, we note that almost all of the symmetric geometries exhibit excellent convexity. Motivated from these considerations, we give several basic selections, e.g., polytope, ellipsoidal and box, of the uncertainty set in the following sections and then derive the robust counterparts of the chance constraint based on the corresponding uncertainty sets.

\subsection{Polytope Model}
The polytope model is parameterized as
\begin{eqnarray}
\mathcal{P}=\{\mathbf{g} | \mid g_{d}-\bar{g}_{d}\mid+\mid g_{c,d}-\bar{g}_{c,d} \mid\leq \Gamma    \},
\end{eqnarray}
where $\bar{g}_{d}, \bar{g}_{c,d}\in \mathbb{R}$ are the center positions of the polytope and $\Gamma>0$ is the budget parameter to control the size of polytope. For learning these parameters, we should collect multiple samples of the uncertain CSI $\mathbf{g}$ as $\mathcal{N}=\{\bm\xi_{1},\bm\xi_{2},\cdots,\bm\xi_{N}\}$, where $\bm\xi_{i}\in \mathbb{R}^{2}$. Then, we propose a statistical learning approach, which includes shape learning and size calibration, for determining the shape parameters.

\subsubsection{Shape Learning} Note that $\bar{\mathbf{g}}=[\bar{g}_{d}, \bar{g}_{c,d}]^{T}$ works as the origin of the polytope uncertainty set. Thus, without loss of generality, it can be chosen as the sample mean, i.e.
\begin{eqnarray}
\bar{\mathbf{g}}=\frac{1}{N} \sum_{k=1}^{N} \bm\xi_{k}.
\end{eqnarray}

After the shape learning, we should determine the size of polytope in the following procedure.

\subsubsection{Size Calibration} The objective of size calibration is to calibrate the uncertainty set so that it can satisfy $\mathbf{g} \in \mathcal{P}(\mathcal{N})$ with confidence $1-\epsilon$. The key idea is to estimate the quantile of the transformation of the data sample. More concretely, let
\begin{eqnarray}
t_{p}(\bm\xi)=\mid \bm\xi(1)-\bar{g}_{d}\mid+\mid \bm\xi(2)-\bar{g}_{c,d} \mid
\end{eqnarray}
be the transformation map from random space $\mathbb{R}^{2}$ into $\mathbb{R}$. Then, the size of $\mathcal{P}$ can be set as the estimated $(1-\epsilon)$-quantile of the underlying distribution of $t_{p}(\bm\xi)$ based on the sample data set $\mathcal{N}$. For achieving this objective, we should introduce the definition of $(1-\epsilon)$-quantile $q_{1-\epsilon}$ as
\begin{eqnarray}
\Pr\left\{t_{p}(\bm\xi)\leq q_{1-\epsilon}\right\}=1-\epsilon.
\end{eqnarray}
Then, by computing the function values of $t_{p}(\bm\xi)$ on all of the samples in $\mathcal{N}$, we can obtain the observations $t_{p}(\bm\xi_{(1)}),\cdots,t_{p}(\bm\xi_{(N)})$. By sorting the observations $t_{p}^{(1)}\leq\cdots\leq t_{p}^{(N)}$ in ascending order, the $k_{p}^{*}=\lceil(1-\epsilon)N\rceil$-th value can be considered as the upper bound of $(1-\epsilon)$-quantile of $t_{p}(\bm\xi)$. As a result, the size of uncertainty set $\mathcal{P}$ can be set as
\begin{eqnarray}
\Gamma=t_{p}(\bm\xi_{(k_{p}^{*})}).
\end{eqnarray}

Using the presented two procedures, the polytope model for covering the uncertain CSI with confidence level $1-\epsilon$ can be obtained. Moreover, we should note that the polytope uncertainty set can be also represented as $ \mathcal{P}=\{\mathbf{g} | \mathbf{M}_{p} (\mathbf{g}-\bar{\mathbf{g}})\leq \bm\Gamma \}$, where $\bm\Gamma=\Gamma\times \mathbf{1}_{4\times1}$, $\mathbf{M}_{p}\in \mathbb{R}^{4\times2}$ is the weight matrix  and each row of $\mathbf{M}_{p}$ indicates a possible realization of the combination of $1$ and $-1$.

Based on the obtained polytope model, the robust counterpart of D2D QoS constraint can be computed as
\begin{eqnarray}
\min &&  \mathbf{p}^{T}\mathbf{g} \\
\textrm{s.t.} && \mathbf{M}_{p} (\mathbf{g}-\bar{\mathbf{g}})\leq \bm\Gamma. \nonumber
\end{eqnarray}
Since (13) is feasible, the optimal objective $\Delta^{*}$ can be obtained via its Lagrange dual as
\begin{eqnarray}
\Delta^{*}=\max &&   -(\bm\Gamma+\mathbf{M}_{p}\bar{\mathbf{g}})^{T} \mathbf{x}     \\
\textrm{s.t.} && -\mathbf{M}_{p}^{T} \mathbf{x} \leq \mathbf{p},\mathbf{x}\geq 0, \nonumber
\end{eqnarray}
where $\mathbf{x}\in \mathbb{R}^{4}$ is the dual variable. If the feasible solution, denoted as $\tilde{\mathbf{x}}$, of (14) satisfy $-(\bm\Gamma+\mathbf{M}_{p}\bar{\mathbf{g}})^{T}\tilde{\mathbf{x}} \geq \gamma_{min}^{d}$, then $\Delta^{*}\geq-(\bm\Gamma+\mathbf{M}_{p}\bar{\mathbf{g}})^{T}\tilde{\mathbf{x}} \geq \gamma_{min}^{d}$ holds for all uncertainty of $\mathbf{g}$. In this case, the D2D QoS constraint in (6c) can be replaced by the following constraint
\begin{eqnarray}
\left\{\begin{array}{l}
-(\bm\Gamma+\mathbf{M}_{p}\bar{\mathbf{g}})^{T} \mathbf{x}  \geq \gamma^{d}_{min}, \\
-\mathbf{M}_{p}^{T} \mathbf{x} \leq \mathbf{p},\mathbf{x}\geq 0.
\end{array}\right.
\end{eqnarray}
It shows that the intractable chance constraint is transformed into a combination of linear constraint. Definitely, it would provide computational advantages for the power allocation problem.

\subsection{Ellipsoidal Uncertainty Set}

The ellipsoidal uncertainty set can be parameterized as
\begin{eqnarray}
\mathcal{E}=\{\mathbf{g}: (\mathbf{g}-\bar{\mathbf{g}})^{T} (\mathbf{g}-\bar{\mathbf{g}})\leq \Lambda \},
\end{eqnarray}
where $\bar{\mathbf{g}}$ is the center of $\mathcal{E}$ and $\Lambda>0$ is the size of $\mathcal{E}$. In the ellipsoidal uncertainty set, the center is same to the one in polytope. The size of $\mathcal{E}$ can be estimated by using the same quantile method as shown in Subsection A. Therefore, we use the following function
\begin{eqnarray}
t_{e}(\bm\xi)=(\bm\xi-\bar{\mathbf{g}})^{T}(\bm\xi-\bar{\mathbf{g}})
\end{eqnarray}
to represent the map from random space $\mathbb{R}^{2}$ into $\mathbb{R}$. By computing the function values of $t_{e}(\bm\xi)$ on each sample of $\mathcal{N}$ and sorting the observations of $t_{e}(\bm\xi)$ in ascending order, the $k_{e}^{*}=\lceil(1-\delta)N\rceil$-th value of the ranked observations can be set as the size of uncertainty set $\mathcal{E}$, i.e.,
\begin{eqnarray}
\Lambda=t_{e}(\bm\xi_{(k_{e}^{*} )}).
\end{eqnarray}

According to the obtained budget parameters, the ellipsoidal uncertainty set can be also represented as
\begin{eqnarray}
\mathcal{E}=\{\mathbf{g}:\mathbf{g}=\bar{\mathbf{g}}+\zeta \mathbf{u},\mathbf{u}^{T}\mathbf{u}\leq1\},
\end{eqnarray}
where $\mathbf{u}\in\mathbb{R}^{2}$ and $\zeta$ can be computed as $\zeta=\sqrt{\Lambda}$.

Based on the ellipsoid uncertainty set, the robust counterpart of D2D QoS constraint can be computed as
\begin{eqnarray}
\min &&  \mathbf{p}^{T}\mathbf{g} \\
\textrm{s.t.} && \mathbf{g}=\bar{\mathbf{g}}+\zeta \mathbf{u},\mathbf{u}^{T}\mathbf{u}\leq 1. \nonumber
\end{eqnarray}
Since
\begin{eqnarray}
\inf_{\parallel\mathbf{u}\parallel\leq 1}  \mathbf{p}^{T}(\bar{\mathbf{g}}+\zeta \mathbf{u})=\mathbf{p}^{T}\bar{\mathbf{g}}-\zeta\parallel\mathbf{p}^{T} \parallel,
\end{eqnarray}
where the derivation is based on the Schwartz inequality, the D2D QoS constraint in (6c) can be replaced by
\begin{eqnarray}
\mathbf{p}^{T}\bar{\mathbf{g}}-\zeta\|\mathbf{p}^{T} \|  \geq \gamma^{d}_{min},
\end{eqnarray}
which is a second-order cone. Thus, it is effectively compatible with the convex optimization tools.

\subsection{Box Uncertainty Set}

The box uncertainty set is parameterized as
\begin{eqnarray}
\mathcal{B}=\{\mid \mathbf{g}(i)-\bar{\mathbf{g}}(i) \mid\leq \Psi, i=1,2\},
\end{eqnarray}
where $\bar{\mathbf{g}}$ is the center of box and $\Psi$ determines the size of $\mathcal{B}$. In order to obtain $\Psi$, we should define
\begin{eqnarray}
t_{b}(\bm\xi_{i})=\max_{i=1,2} \mid \bm\xi_{i}-\bar{\mathbf{g}}_{i} \mid
\end{eqnarray}
to map the channel samples into real number. Then, the parameter $\Psi$ can be chosen as the $k_{b}^{*}=\lceil(1-\epsilon)N\rceil$-th value of the ranked observations $t_{b}^{(1)}\leq\cdots\leq t_{b}^{(N)}$ in ascending order, i.e., $\Psi=t_{b}(\bm\xi_{(k_{b}^{*})})$.

Then, based on the obtained parameter, the box uncertainty set can be re-represented as
\begin{eqnarray}
\mathcal{B}=\{\mathbf{g} | \mathbf{M}_{b} (\mathbf{g}-\bar{\mathbf{g}})\leq \bm\Psi \},
\end{eqnarray}
where $\bm\Psi=\Psi\times \mathbf{1}_{4\times1}$ and each row of $\mathbf{M}_{b}$ indicates a possible realization of the combination $0$ and $\pm1$.

Given this uncertainty set, the robust counterpart can be computed as
\begin{eqnarray}
\textrm{min} &&  \mathbf{p}^{T}\mathbf{g} \\
\textrm{s.t.} && \mathbf{M}_{b} (\mathbf{g}-\bar{\mathbf{g}})\leq \bm\Psi.  \nonumber
\end{eqnarray}
The optimal objective $\Theta^{*}$ of problem (26) can be obtained via its Lagrange dual as
\begin{eqnarray}
\Theta^{*}=\textrm{max} &&   -(\bm\Psi+\mathbf{M}_{b}\bar{\mathbf{g}})^{T} \mathbf{y}     \\
\textrm{s.t.} && -\mathbf{M}_{b}^{T} \mathbf{y} \leq \mathbf{p},\mathbf{y}\geq 0, \nonumber
\end{eqnarray}
where $\mathbf{y}\in \mathbb{R}^{4}$ is the dual variable. Similar to the computation of robust counterpart based on polytope uncertainty set, we obtain  $\Theta^{*}\geq-(\bm\Psi+\mathbf{M}_{b}\bar{\mathbf{g}})^{T}\tilde{\mathbf{y}} \geq \gamma_{min}^{d}$ for all uncertainties of $\mathbf{g}$. Finally, the chance constraint of D2D can be replaced by the following equations
\begin{eqnarray}
\left\{\begin{array}{l}
-(\bm\Psi+\mathbf{M}_{b}\bar{\mathbf{g}})^{T} \mathbf{y}  \geq \gamma^{d}_{min}, \\
-\mathbf{M}_{b}^{T} \mathbf{y} \leq \mathbf{p},\mathbf{y}\geq 0.
\end{array}\right.
\end{eqnarray}
It shows that the robust counterpart of chance constraint supported on box uncertainty set leads to a combination of linear constraint. It is definitely computationally convenient for the convex optimization tools.

\subsection{Power Allocation}

By modeling the uncertain CSI, the intractable chance QoS constraint of D2D can be transformed into the corresponding tractable one. Then, power allocation problem (6) can be reformulated as the following optimization
\begin{eqnarray}
\max\limits_{p_{c},p_{d}} && B\log_{2}\left(1+\frac{p_{c}g_{c}}{\sigma^{2}+p_{d}g_{d,B}}\right) \\
\textrm{s.t.} &&  \textrm{(15)} \ \textrm{or} \ \textrm{(22)} \ \textrm{or} \ \textrm{(28)}  \nonumber \\
&&\textrm{(6b)}, \textrm{(6d)},\textrm{(6e)}. \nonumber
\end{eqnarray}
Although any combination of the constraints in (29) can constitutes a convex set, problem (29) is still nonconvex due to the fractional form of $p_{c}$ and $p_{d}$ in objective function. For solving this nonconvex problem, we should first fix either $p_{c}$ or $p_{d}$ to find the corresponding optimal $p_{d}^{*}$ or $p_{c}^{*}$, respectively. Then, repeat the above process until both $p_{c}$ and $p_{d}$ converge the optimal solution of (29).

Inspired by the above discussion, we can first consider a fixed $p_{d}$ and formulate a subproblem to maximize the throughput of CUE, i.e.
\begin{eqnarray}
\max\limits_{p_{c}}&&   B\log_{2}\left(1+\frac{p_{c}g_{c}}{\sigma^{2}+p_{d}g_{d,B}}\right) \\
\textrm{s.t.}&& \textrm{(15)} \ \textrm{or} \ \textrm{(22)} \ \textrm{or} \ \textrm{(28)}  \nonumber \\
&& p_{c}\geq0, \textrm{(6b)},  \nonumber
\end{eqnarray}
which is a concave optimization problem. Thus, it can be effectively solved by the widely used convex optimization tools. Then, before developing the algorithm for searching $p_{d}$, we should give the following observations.

For the power allocation in D2D underlaying cellular network, if the CUE transmit power is larger, more transmit power need to be allocated for satisfying the QoS constraint of D2D. Therefore, $p_{d}(p_{c})$ can be considered as an increasing function of $p_{c}$. Based on this observation, the optimal transmit power of DUE can be searched by using the bisection search algorithm. Moreover, the following theorem provides a condition to terminate the searching process.

\begin{lemma}
The optimal power allocation solution for (29) must satisfy either $p_{c}^{*}=P_{max}^{c}$ or $p_{d}^{*}=P_{max}^{d}$.
\end{lemma}
\proof The similar proof can be found in \cite{C.GuoJSAC}.
\endproof

Based on \textbf{Lemma 1}, the bisection search method can be developed in \textbf{Algorithm 1} for searching the optimal power allocation of (29). Then, the standard operating procedure for the symmetrical-geometry-based robust resource allocation can be concluded as follows:
\begin{itemize}
  \item Collect a set $\mathcal{N}$ of samples of the imperfect CSI.
  \item Learn the shape parameter $\bar{\mathbf{g}}$ based on the sample set $\mathcal{N}$.
  \item Set the size parameter based on quantize estimation method.
  \item Construct the robust counterpart based on the obtained uncertainty set, and use it to replace the chance constraint in (5c) to derive the convex constraint set.
  \item Solve the induced power allocation problem by using \textbf{Algorithm 1}.
\end{itemize}

\begin{algorithm}[h]\label{Am1}
\caption{Bisection Search-Based Power Allocation}
\begin{algorithmic}[0]
\STATE Set termination threshold $0<\zeta<1$;
\STATE Set $p_{k,min}^{d}=0$ and $p_{k,max}^{d}=P_{max}^{d}$;
\WHILE {$p_{k}^{d}<P_{max}^{d}-\zeta$}
\STATE set $p_{k}^{d}=(p_{k,min}^{d}+p_{k,max}^{d})/2$; Solve (30) to obtain $p_{m}^{c}$;
\IF{$p_{m}^{c}>P_{max}^{c}+\zeta$}
\STATE $p_{k,max}^{d}=p_{k}^{d}$
\ELSIF{$p_{m}^{c}<P_{max}^{c}-\zeta$}
\STATE $p_{k,min}^{d}=p_{k}^{d}$
\ELSIF{$P_{max}^{c}-\zeta<p_{m}^{c}<P_{max}^{c}+\zeta$}
\STATE break
\ENDIF
\ENDWHILE
\STATE Output the optimal transmit powers $p_{m}^{c,*}$ and $p_{k}^{d,*}$.
\end{algorithmic}
\end{algorithm}

\section{Kernel Learning Induced Robust Resource Allocation}

In above section, all of the uncertainty sets are symmetric and radial. The resulting robust counterparts have the succinct formulations and are computational convenience for the resource allocation problem. However, we should note that the classical uncertainty sets have some limitations in practical applications. First, we should determine the moment values as well as the support information carefully. As illustrated in the channel training process, it is a nontrivial task to obtain the exact information of these parameters. Second, the CSI uncertainties on different dimensions in D2D underlaying cellular network maybe intertwined and asymmetric, as shown in Fig. 2(a). No matter what the shape of the symmetric-geometry-based uncertainty set is, it cannot accurately cover the distributions of the uncertain CSI. As a result, the symmetric-geometry-based uncertainty sets lead to large block superfluous coverage so that they will result in over-conservative decisions. Inspired by the recently popular machine learning (ML) methods, constructing the uncertainty set can be considered as the unsupervised leaning problem. The support vector clustering (SVC) technique in ML is always considered as an effective learning method to provide powerful representations of sample distribution \cite{2017Data,2001Support}. Thus, in this section, we propose to resort the SVC technique to estimate the uncertainty set from  CSI samples.

\begin{figure}
\begin{center}
\includegraphics[width=3.6in,height=1.9in]{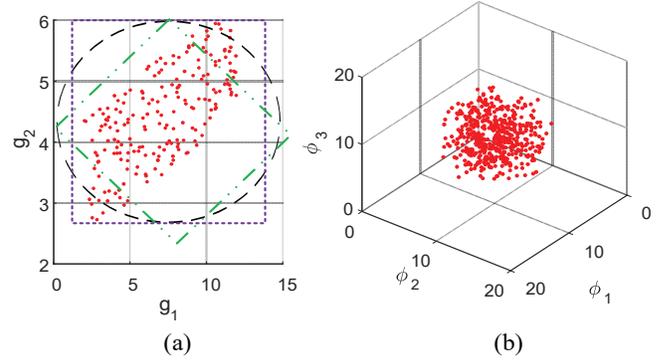}
\label{fig2}\caption{Uncertainty set construction based on the imperfect CSI samples. (a) Symmetric-geometry-based uncertainty sets, i.e., polytope, ellipsoidal and box. (b) Nonlinear mapping map of the data samples into a high-dimensional features space.}
\end{center}
\end{figure}

%Recalled the statements in (33), $\epsilon$ is the parameter to control the size of the sphere in high-dimensional features space. Namely, the obtained sphere will at least encapsulate $(1-\epsilon)\times 100\%$ percentage of the training samples. Thus, there are roughly $\epsilon N$ samples located in the exterior of the sphere, i.e., $\|\phi(\bm\xi_{i})-\bm\rho\|^{2}\geq R^{2}$, which is referred to as support supports. Therefore, the number of  auxiliary variables, i.e. $\bm\upsilon_{i}$ and $\bm\omega_{i}$, is at least $\lceil2N\epsilon\rceil$. This indicates that the parameter $\epsilon$ is also can be used to manipulate the complexity of the induced robust optimization.

\subsection{SVC-Based Robust Optimization}

It has been widely proven that the SVC is an efficient pattern recognition approach for analyzing the complicated high-dimensional data. The SVC focuses on the description of channel uncertainties by means of an enclosing sphere with minimal volume. To achieve this objective, we should first collect a set of $\mathcal{N}=\{\bm\xi_{1},\bm\xi_{2},\cdots,\bm\xi_{N}\}$ samples of the imperfect CSI and then use a nonlinear mapping $ \phi(\bm\xi_{i}):\mathbb{R}^{2}\mapsto \mathbb{R}^{K}$ to map the CSI samples into a high-dimensional features space $\mathcal{F}$, i.e., $K>2$. After the nonlinear transformation, the CSI samples with correlation and asymmetry can be gathered into cluster in the high-dimensional features space, as shown in Fig. 2(b). The objective of SVC is to seek the smallest sphere to enclose the CSI samples in high-dimensional features space, which can be formulated as the following optimization problem.
\begin{subequations}
\begin{eqnarray}
\min\limits_{R,\bm\rho,\{\psi_{i}\}}&&\!\!\!\!\!\!\!\!\!  R^{2}+C \sum_{i=1}^{N} \psi_{i}  \\
\textrm{s.t.} &&\!\!\!\!\!\!\!\!\!  \| \phi(\bm\xi_{i})-\bm\rho \|^{2}  \leq R^{2}+ \psi_{i}, i=1,\cdots,N,  \\
&&\!\!\!\!\!\!\!\!\!  \psi_{i}\geq0, i=1,\cdots,N,
\end{eqnarray}
\end{subequations}
where $R$ denotes the radius of the sphere, $\bm\rho$ is the center, $\psi_{i}$ is a slack variable and $C$ is the parameter to control the tradeoff between the two goals.

%In other words, the obtained high-dimensional features space will at least encapsulate $(1-\epsilon)\times 100\%$ percentage of the $N$ training samples. At this case, we denote the obtained uncertainty set as $\mathcal{S}_{\epsilon}(N)$, which captures the feasible region with  $(1-\epsilon)\times 100\%$ confidence.

We introduce the Lagrangian multipliers $\bm\lambda$ and $\bm\eta$ for constructing the following Lagrangian function
\begin{eqnarray}
L(R,\bm\rho,\bm\psi,\bm\lambda,\bm\eta)=\!\!\!\!\!\!\!\!\!&&R^{2}+C\sum_{i=1}^{N}\psi_{i}-\sum_{i=1}^{N}\eta_{i}\psi_{i}\\
&&+\sum_{i=1}^{N}\lambda_{i}(\| \phi(\bm\xi_{i})-\bm\rho\|^{2} - R^{2}-\psi_{i}). \nonumber
\end{eqnarray}
The derivatives of $L$ with respect to $R$, $\bm\rho$ and $\psi_{i}$ can be computed as
\begin{eqnarray}
\left\{\begin{array}{l}
         \frac{\partial L}{\partial R}=0 \rightarrow   \sum_{i=1}^{N}\lambda_{i}=1, \\
         \frac{\partial L}{\partial\bm\rho}=\mathbf{0}\rightarrow \bm\rho=\sum_{i=1}^{N}\lambda_{i}\bm\phi(\bm\xi_{i}),\\
         \frac{\partial L}{\partial\psi_{i}}\rightarrow \lambda_{i}+\eta_{i}=C.
       \end{array}
 \right.
\end{eqnarray}
Eq. (33) shows that the center of sphere is a linear combination of the mapping of all uncertain channel samples. Moreover, according to the Karuch-Kuhn-Tucker (KKT) conditions, one can obtain
\begin{eqnarray}
\eta_{i} \psi_{i}=0,\ \lambda_{i} (\| \phi(\bm\xi_{i})-\bm\rho\|^{2} - R^{2}-\psi_{i})=0.
\end{eqnarray}
Using these relations, we may eliminate the variables $R$, $\bm\rho$ and $\psi_{i}$,  and turn the Lagrangian into a disciplined quadratic programming (QP) as the dual problem
\begin{subequations}
\begin{eqnarray}
&&\min\limits_{\bm\lambda} \sum_{i=1}^{N}\sum_{j=1}^{N}\lambda_{i}\lambda_{j}K(\bm\xi_{i},\bm\xi_{j})-\sum_{i=1}^{N}\lambda_{i}K(\bm\xi_{i},\bm\xi_{i})   \\
&& \ \textrm{s.t.} \quad  0\leq \lambda_{i}\leq C,i=1,\cdots N,   \\
&&\quad\quad \  \sum_{i=1}^{N}\lambda_{i}=1,
\end{eqnarray}
\end{subequations}
where $K(\cdot,\cdot)$ is the kernel trick and it can be computed as  $K(\bm\xi_{i},\bm\xi_{j})=\bm\phi(\bm\xi_{i})^{T}\bm\phi(\bm\xi_{j})$. We should note that the specific expression of kernel function should not affect the convexity of the dual problem. In the following discussions, we will illustrate how to design the kernel in detail.

Based on the KKT conditions and complementary slackness, we can obtain the following desirable geometric interpretations,
\begin{eqnarray}
\left\{\begin{array}{l}
         \|\phi(\bm\xi_{i})-\bm\rho \|^{2}< R^{2}\rightarrow \lambda_{i}=0, \eta_{i}=C, \\
         \|\phi(\bm\xi_{i})-\bm\rho\|^{2}=R^{2}\rightarrow 0<\lambda_{i}<C,0<\eta_{i}<C, \\
         \|\phi(\bm\xi_{i})-\bm\rho\|^{2}> R^{2}\rightarrow \lambda_{i}=C, \eta_{i}=0.
       \end{array}
 \right.
\end{eqnarray}
It is not difficult to observe that the CSI samples located in the interior of sphere has no contribution on the construction of the center. We denote the set of these samples as $\mathcal{I}$. Thus, the uncertain CSI samples with positive $\lambda_{i}$ can be referred to as support vectors. Among the supports, there are some CSI samples $\bm\xi_{i}$ with $0<\lambda_{i}<C$ located exactly on the boundary of the sphere. We call these CSI samples as boundary support vectors. The rest of them with $\lambda_{i}=C$ are regarded as outliers. Based on these observations, we give the following definitions
\begin{eqnarray}
\mathcal{F}=\{i\mid \lambda_{i}>0, \forall i \} \ \textrm{and} \ \mathcal{B}_{v}=\{i\mid 0<\lambda_{i}<C, \forall i \}
\end{eqnarray}
to denote the index sets of all support vectors and boundary support vectors, respectively. Then, based on (33) and (37), we obtain
\begin{eqnarray}
1=\sum_{i\in \mathcal{I}} \lambda_{i}+\sum_{i\in \mathcal{F}-\mathcal{B}_{v}}\lambda_{i}+\sum_{i\in \mathcal{B}_{v}}\lambda_{i}>n_{out}C,
\end{eqnarray}
where $n_{out}$ denote the number of outlier samples. Then, $1/(NC)$ can be considered as the upper bound on the fraction of outliers.  For satisfying the chance constraint of D2D communication, the objective of (31) is to at least encapsulate $(1-\epsilon)\times 100\%$ percentage of the $N$ CSI samples. At this case, the fraction of outliers should be smaller than $\epsilon\times 100\%$. Therefore, it is naturally to set $C=1/(\epsilon N)$ to control the sphere covering the CSI samples with  $(1-\epsilon)\times 100\%$ confidence.

Then, the radius of the sphere can be determined as the distance from the center $\bm\rho$ to any boundary support vector $\bm\xi_{l}$ in $\mathcal{B}_{v}$, i.e.,
\begin{eqnarray}
R^{2}\!\!\!\!\!\!\!\!\!\!\!&&=\|\phi(\bm\xi_{l})-\bm\rho\|^{2}=\|\phi(\bm\xi_{l})-\sum_{i=1}^{N}\lambda_{i}\bm\phi(\bm\xi_{i})\|^{2}\\
&&\!\!\!\!=K(\bm\xi_{l},\bm\xi_{l})\!-\!2\sum_{i=1}^{N}\lambda_{i}K(\bm\xi_{l},\bm\xi_{i})\!+\!\sum_{i=1}^{N}\sum_{j=1}^{N}\lambda_{i}\lambda_{j}K(\bm\xi_{i},\bm\xi_{j}). \nonumber
\end{eqnarray}
Thus, the feasible set of uncertain CSI can be defined as the sphere with radius $R^{2}$, i.e.
\begin{eqnarray}
\mathcal{S}_{\epsilon}(\mathcal{N})=&&\!\!\!\!\!\!\left\{\mathbf{g}\mid K(\mathbf{g},\mathbf{g})-2\sum_{i=1}^{N} \lambda_{i}K(\mathbf{g},\bm\xi_{i})\right.\\
&&\quad\quad\quad\left.+\sum_{i=1}^{N}\sum_{j=1}^{N}\lambda_{i}\lambda_{j}K(\bm\xi_{i},\bm\xi_{j})\leq R^{2} \right\}. \nonumber
\end{eqnarray}
Indeed, the feasible set $\mathcal{S}_{\epsilon}(\mathcal{N})$ can be recognized a data-driven uncertainty set. The kernel function in $\mathcal{S}_{\epsilon}(\mathcal{N})$ plays a key role for ensuring the application of $\mathcal{S}_{\epsilon}(\mathcal{N})$ in robust optimization. The commonly used kernel function, such as the Gaussian kernel: $K(\bm\xi_{i},\bm\xi_{j})=\exp(-q\|\bm\xi_{i}-\bm\xi_{j}\|^{2})$, sigmoid kernel: $K(\bm\xi_{i},\bm\xi_{j})=\tanh(a\cdot\bm\xi_{i}^{T}\cdot\bm\xi_{j}+r)$ and polynomial kernel: $K(\bm\xi_{i},\bm\xi_{j})=(\bm\xi_{i}^{T}\cdot\bm\xi_{j}+1)^{d}$ contains some nonlinear terms \cite{2004Fuzzy}, which inevitably complicate its application in power allocation. Hence, when designing the kernel function, it is imperative to consider that the designed kernel function could provide a convex expression of $\mathcal{S}_{\epsilon}(\mathcal{N})$  and preserve the convexity of the dual problem in (35). Based on these motivations, the weighted generalized intersection kernel (WGIK) \cite{2017Data} is employed for the robust optimization
\begin{eqnarray}
K(\bm\xi_{i},\bm\xi_{j})=\sum_{k=1}^{2}\Xi_{k}-\|\mathbf{Q}(\bm\xi_{i}-\bm\xi_{j}) \|_{1},
\end{eqnarray}
where $\mathbf{Q} \in \mathbb{R}^{2\times 2}$ is a weighted matrix, $\Xi_{k}$ represents the interval width. In these parameters, $\mathbf{Q}$ can be constructed as $\mathbf{Q}=\bm\Sigma^{-\frac{1}{2}}$, where $\bm\Sigma$ is the covariance information from the uncertain CSI samples and it is computed as
\begin{eqnarray}
\bm\Sigma\!=\!\frac{1}{N\!-\!1}\left[\sum_{i=1}^{N\!-\!1}\bm\xi_{i}\bm\xi_{i}^{T}\!-\!\frac{1}{N\!-\!1}\left(\sum_{i=1}^{N}\bm\xi_{i}\right)\left(\sum_{i=1}^{N}\bm\xi_{i}\right)^{T} \right].
\end{eqnarray}

Since the dual problem in (35) is convex only when the kernel matrix $\mathbf{K}=\{K(\bm\xi_{i},\bm\xi_{j})\}\succ \mathbf{0}$, $\Xi_{k}$ is chosen with a simple criterion as
\begin{eqnarray}
\Pi_{k}>\max_{1\leq i\leq N} \mathbf{q}_{k}^{T}\bm\xi_{i}-\min_{1\leq i\leq N}\mathbf{q}_{k}^{T}\bm\xi_{i}.
\end{eqnarray}
where $\mathbf{q}_{k}$ is the column vector of $\mathbf{Q}$. Based on the WGIK model, we notice that $K(\bm\xi_{i},\bm\xi_{i})$ is a constant for any $i$ which satisfies $1\leq i \leq N$.

By substituting (39) into (40), the uncertainty set can be further computed as follows
\begin{eqnarray}
\mathcal{S}_{\epsilon}(\mathcal{N})\!=\!\left\{\!\mathbf{g}\!\mid \sum_{i=1}^{N}\lambda_{i}K(\mathbf{g},\bm\xi_{i})\!\geq\!  \sum_{i=1}^{N}\lambda_{i}K(\bm\xi_{l},\bm\xi_{i}), l\!\in\!\mathcal{B}_{v}  \right\}.
\end{eqnarray}
Recalled the discussion in Eq. (36), when sample $i$ located in the interior of the sphere, there exists $\lambda_{i}=0$. It indicates that only the CSI samples in the exterior of the sphere can contribute to the construction of uncertainty set. Thus, (44) can be further simplified as
\begin{eqnarray}
\mathcal{S}_{\epsilon}(\mathcal{N})\!=\!\left\{\!\mathbf{g}\!\mid \sum_{i\in \mathcal{F}}\lambda_{i}K(\mathbf{g},\bm\xi_{i})\!\geq\!  \sum_{i\in \mathcal{F}}\lambda_{i}K(\bm\xi_{l},\bm\xi_{i}), l\!\in\!\mathcal{B}_{v}  \right\}.
\end{eqnarray}
Then, by substituting Eq. (41) into Eq. (45), the explicit expression of data-driven uncertainty set is represented as
\begin{eqnarray}
\mathcal{S}_{\epsilon}(\mathcal{N})\!\!\!\!\!\!\!\!\!\!&&=\\
&&\!\!\!\!\!\!\!\!\!\!\left\{\!\mathbf{g}\!\mid \! \sum_{i\in\mathcal{F} }\!\lambda_{i}\|\!\mathbf{Q}(\mathbf{g}\!-\!\bm\xi_{i})\|_{1}\!\!\leq\!\! \sum_{i\in\mathcal{F}}\!\lambda_{i}\|\!\mathbf{Q}(\bm\xi_{l}\!\!-\!\bm\xi_{i})\|_{1}, l\!\in \!\mathcal{B}_{v}  \right\}. \nonumber
\end{eqnarray}
It is not difficult to observe that the right hand side of (46) is a constant. Then, let $\varrho=\sum_{i\in\mathcal{F}}\lambda_{i} \|\mathbf{Q}(\bm\xi_{l}-\bm\xi_{i})\|_{1}, l\in\mathcal{B}_{v}$ and introduce auxiliary variable $\mathbf{V}=[\mathbf{v}_{1},\mathbf{v}_{1},\cdots,\mathbf{v}_{N}]$, where $\mathbf{v}_{i}\in \mathbb{R}^{2}$, the uncertainty set $\mathcal{S}_{v}(\mathcal{N})$ can be further rewritten as
\begin{eqnarray}
\mathcal{S}_{\epsilon}(\mathcal{N})=\left\{\mathbf{g}\left|\begin{array}{c}
                                       \exists \mathbf{v}_{i}, i\in\mathcal{F} \ \textrm{s.t.} \\
                                       \sum_{i\in\mathcal{F}}(\lambda_{i} \cdot\mathbf{v}_{i}^{T}\mathbf{1}_{2\times1})\leq \varrho \\
                                       -\mathbf{v}_{i}\leq \mathbf{Q} (\mathbf{g}-\bm\xi_{i})\leq \mathbf{v}_{i}, i\in \mathcal{F}
                                     \end{array}
\right.\right\}.
\end{eqnarray}

\begin{figure}
\begin{center}
\includegraphics[width=3.3in,height=1.8in]{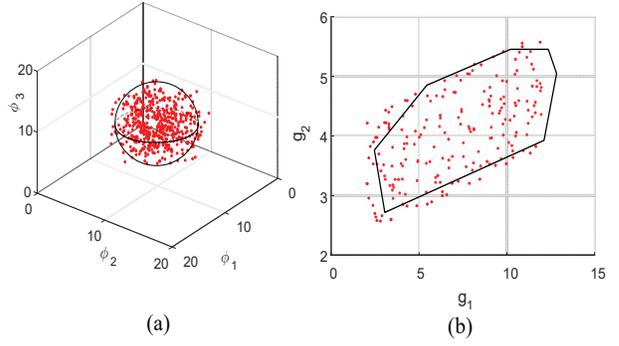}
\label{fig2}\caption{Transformation of uncertainty set from high-dimensional feature space to two-dimensional CSI space. (a) Sphere in high-dimensional feature space. (b) Polytope in the two-dimensional CSI space.}
\end{center}
\end{figure}
Eq. (47) shows that the sphere in high-dimensional feature space is transformed into a polytope in two-dimensional CSI space. Hence, it will provide computational advantages for the power allocation problem. This is shown in Fig. 3. The obtained polytope is not symmetric and radial, moreover it can cover the uncertain CSI without surplus. Besides, the constraints $-\mathbf{v}_{i}\leq \mathbf{Q}(\mathbf{g}-\bm\xi_{i})\leq \mathbf{v}_{i}, i\in \mathcal{F}$ constitute the edges of a polytope. The number of edges depends on the number of support vectors in $\mathcal{F}$.

Then, under the polytope uncertain set, the chance QoS constraint of D2D can be rewritten as the following linear programming (LP)
\begin{subequations}
\begin{eqnarray}
\min_{\mathbf{g},\{\mathbf{v}_{i}\}} &&  \mathbf{p}^{T}\mathbf{g} \\
\textrm{s.t.} && \sum_{i\in\mathcal{F}}(\lambda_{i} \cdot\mathbf{v}_{i}^{T}\mathbf{1}_{2\times 1})\leq \varrho,\\
&&  -\mathbf{v}_{i}\leq \mathbf{Q} (\mathbf{g}-\bm\xi_{i})\leq \mathbf{v}_{i}, i\in \mathcal{F}.
\end{eqnarray}
\end{subequations}
By introducing the  Lagrange multipliers $\kappa$, $\bm\varphi_{i}$ and $\bm\omega_{i}$, the dual problem of (48) can be given as
\begin{subequations}
\begin{eqnarray}
\max\limits_{\kappa,\{\bm\varphi_{i}\},\{\bm\omega_{i}\}} && \sum_{i\in \mathcal{F}}(\bm\omega_{i}-\bm\varphi_{i})^{T}\mathbf{Q}\bm\xi_{i}-\varrho\kappa   \\
 \textrm{s.t.} &&\sum_{i\in F}(\bm\omega_{i}-\bm\varphi_{i})^{T}\mathbf{Q}-\mathbf{p}=\mathbf{0}, \\
&& \bm\omega_{i}+\bm\varphi_{i}-\lambda_{i}\cdot\kappa\cdot\mathbf{1}=\mathbf{0},\forall i\in \mathcal{F}, \\
&&\kappa\geq 0,\bm\varphi_{i},\bm\omega_{i}  \in \mathbb{R}^{2}_{+}.
\end{eqnarray}
\end{subequations}
In this case, the intractable chance constraint in (5c) can be replaced by the following linear constraint
\begin{eqnarray}
\left\{\begin{array}{l}
         \sum_{i\in \mathcal{F}}(\bm\omega_{i}-\bm\varphi_{i})^{T}\mathbf{Q}\bm\xi_{i}-\varrho\kappa\geq \gamma_{min}^{d}, \\
         \sum_{i\in \mathcal{F}}(\bm\omega_{i}-\bm\varphi_{i})^{T}\mathbf{Q}-\mathbf{p}=\mathbf{0}, \\
         \bm\omega_{i}+\bm\varphi_{i}-\lambda_{i}\cdot\kappa\cdot\mathbf{1}=\mathbf{0},\forall i\in \mathcal{F}, \\
         \kappa\geq 0,\bm\varphi_{i},\bm\omega_{i}  \in \mathbb{R}^{2}_{+}.
       \end{array}
\right.
\end{eqnarray}
Compared with the chance constraint, the deterministic constraint is tractable but the introduced auxiliary variables will increase the complexity of the problem. In words, the complexity increases with the number of support vectors. Recalled the statements below Eq. (38), $\epsilon$ is the parameter to control the size of the sphere in high-dimensional features space. Namely, the obtained sphere will at least encapsulate $(1-\epsilon)\times 100\%$ percentage of the training samples. Thus, there are roughly $\epsilon N$ samples located in the exterior of the sphere, i.e., $\|\phi(\bm\xi_{i})-\bm\rho\|^{2}\geq R^{2}$, which is referred to as support supports. Therefore, the number of  auxiliary variables, i.e. $\bm\upsilon_{i}$ and $\bm\omega_{i}$, is at least $\lceil2N\epsilon\rceil$. This indicates that the parameter $\epsilon$ is also can be used to manipulate the complexity of the induced robust optimization.

\subsection{Power Allocation}
After clustering the uncertain CSI, the intractable chance QoS constraint of D2D can be transformed into a combination of linear constraints. Then, the power allocation problem can be reformulated as
\begin{eqnarray}
\max\limits_{p_{c},p_{d}} && B\log_{2}\left(1+\frac{p_{c}g_{c}}{\sigma^{2}+p_{d}g_{d,B}}\right) \\
\textrm{s.t.} &&  \textrm{(50)}, \textrm{(5b)}, \textrm{(5d)}, \textrm{(5e)}. \nonumber
\end{eqnarray}
Evidently, the problem in (51) is similar to the one in (29). Thus, the same bisection search method as illustrated in \textbf{Algorithm 1} can be deployed for searching the optimal solution. To sum up, we propose the following standard operating procedure for the SVC-based power allocation.
\begin{itemize}
  \item Collect a set $\mathcal{N}$ of samples of the imperfect CSI;
  \item Calculate the covariance matrix $\bm\Sigma$ based on the CSI sample set $\mathcal{N}$ and obtain the weight matrix $\mathbf{Q}=\bm\Sigma^{\frac{1}{2}}$;
  \item Compute the interval width $\Xi_{k}$ according to the proposed criterion in (43);
  \item Obtain the kernel trick based on the WGIK in (41);
  \item Solve the disciplined QP in (35) for determining the solution $\bm\lambda$ and the indices of support vectors;
  \item Use the support vectors $\{\bm\xi_{i},i\in \mathcal{F}\}$ and the corresponding Lagrange multipliers $\{\lambda_{i},i\in \mathcal{F}\}$ to construct the robust counterpart in (50), and use it to replace the chance constraint in (6c) to derive the convex constraint set.
  \item Solve the induced power allocation problem by using \textbf{Algorithm 1}.
\end{itemize}

\section{Simulation Results}

%the positions and the distances between different nodes are illustrated in Fig. 4.

In this section, we conduct the simulation to verify the performance of the proposed approaches. In the simulation, we consider a single cell simulation model as the one in Fig. 1. The distances between different nodes are set as $D_{c}=42$m, $D_{c,d}=50.16$m, $D_{d,B}=85$m and $D_{d}=44$m, respectively. The carrier bandwidth is set as $B=10$MHz. The large-scale channel gain in the cellular network is composed of shadow fading and path loss, where the shadowing standard deviation is set as 8dB and the path loss is modeled as the macrocell propagation model $128.1+37.6\log_{10}d(\textrm{km})$. On the contrary, the path loss for D2D communication is set as the WINNER+B1 \cite{IST2008WINNER} model and the shadowing standard deviation is 4dB. Besides, all of the small-scale channel gains are modeled as Rayleigh fading and the noise power is set as $-134$dBm. In the simulation, we consider two types of uncertainties under different distributions: (i) exponential distribution truncated on a polytope; (ii) bivariate Gaussian. The sample number for channel training is $N=1000$. The resource allocation solutions are tested in a test set with 10000 samples of the uncertain CSI. The other parameters are given separately for each experiment. We compare our proposed approaches with the Non-Robust policy as a baseline, where the power allocation problem is solved based on the average channel gain $\bar{\mathbf{g}}$. Besides that, we also develop a Quantile-SVC approach in the following as a baseline to illustrate the effectiveness of the proposed SVC-based robust optimization.

%\begin{figure}
%\begin{center}
%\includegraphics[width=3.2in,height=2.6in]{simulation_model.eps}
%\label{fig2}\caption{Simulation model.}
%\end{center}
%\end{figure}

\begin{figure*}[!t]
        \centering
        \subfigure[Symmetrical-geometry-based uncertainty set under truncated exponential.]{
          \begin{minipage}[b]{0.3\linewidth}
            \centering
            \resizebox{\linewidth}{!}{%
              \includegraphics[height=0.5in]%
              {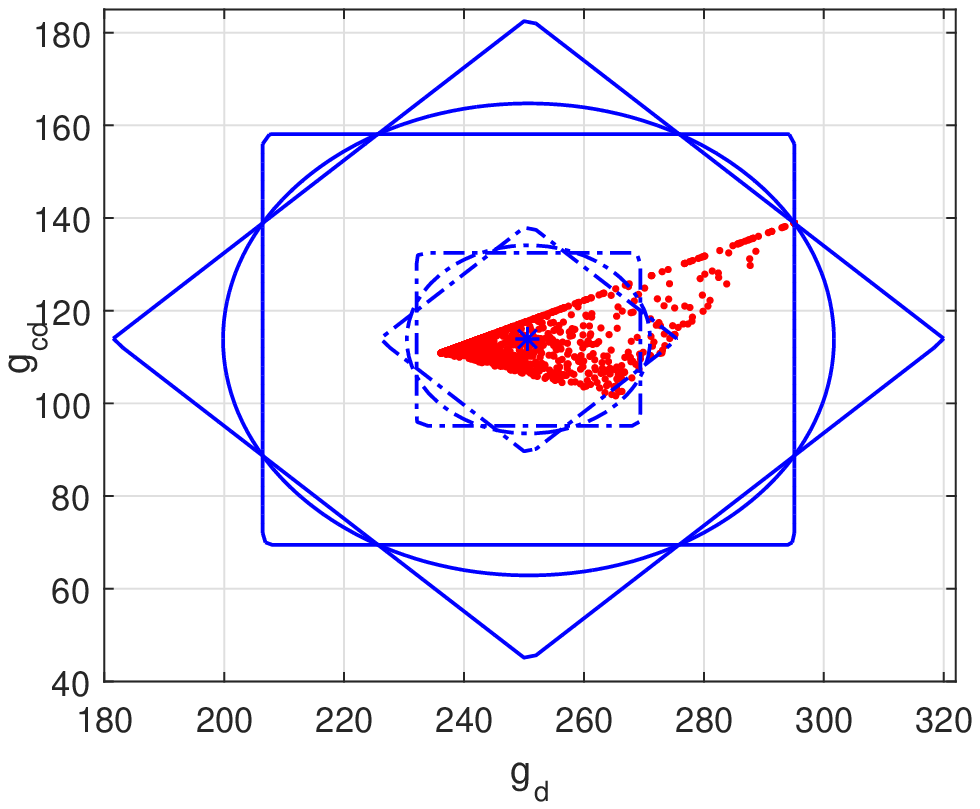}}
            \end{minipage}}
        \subfigure[Symmetrical-geometry-based uncertainty set under gaussian.]{
          \begin{minipage}[b]{0.3\linewidth}
            \centering
            \resizebox{\linewidth}{!}{%
              \includegraphics[height=0.5in]%
              {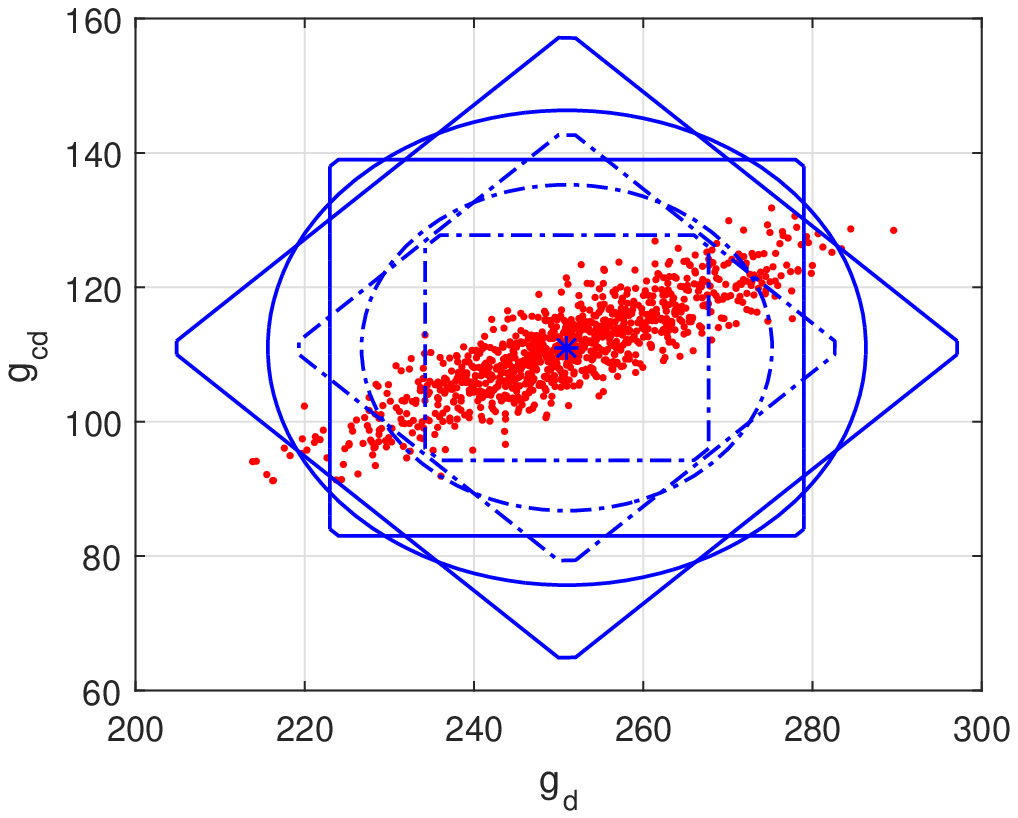}}
            \end{minipage}}
                \subfigure[Quantile-SVC approach under truncated exponential.]{
          \begin{minipage}[b]{0.3\linewidth}
            \centering
            \resizebox{\linewidth}{!}{%
              \includegraphics[height=0.5in]%
              {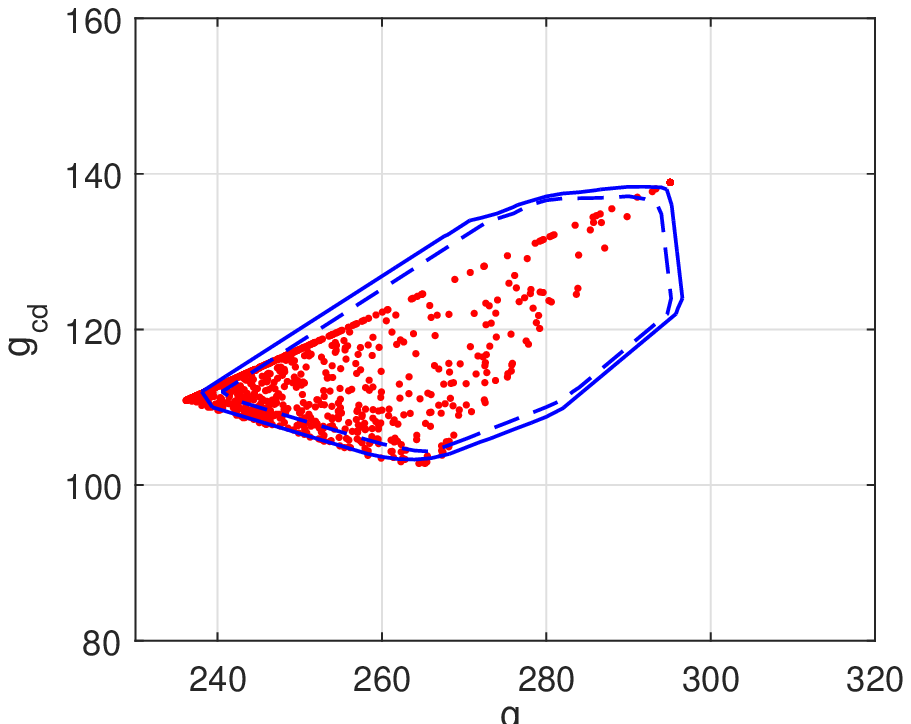}}
            \end{minipage}}
                \subfigure[SVC under truncated exponential.]{
          \begin{minipage}[b]{0.3\linewidth}
            \centering
            \resizebox{\linewidth}{!}{%
              \includegraphics[height=0.5in]%
              {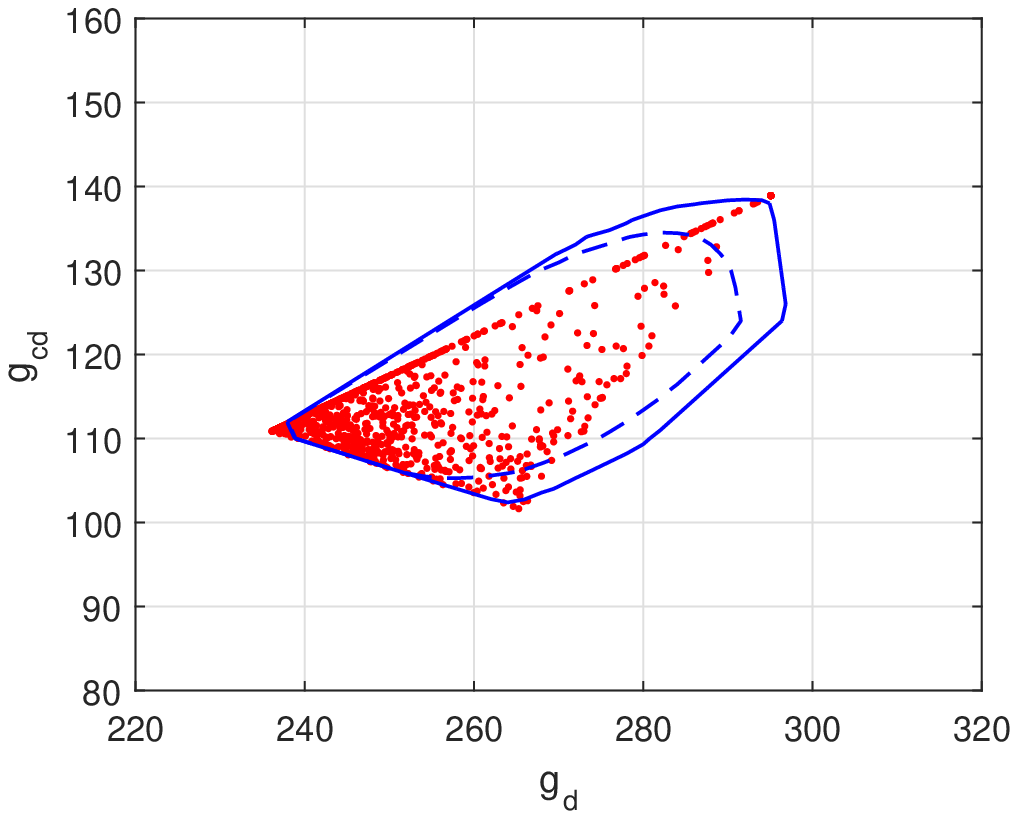}}
            \end{minipage}}
                \subfigure[Quantile-SVC approach under gaussian.]{
          \begin{minipage}[b]{0.3\linewidth}
            \centering
            \resizebox{\linewidth}{!}{%
              \includegraphics[height=0.5in]%
              {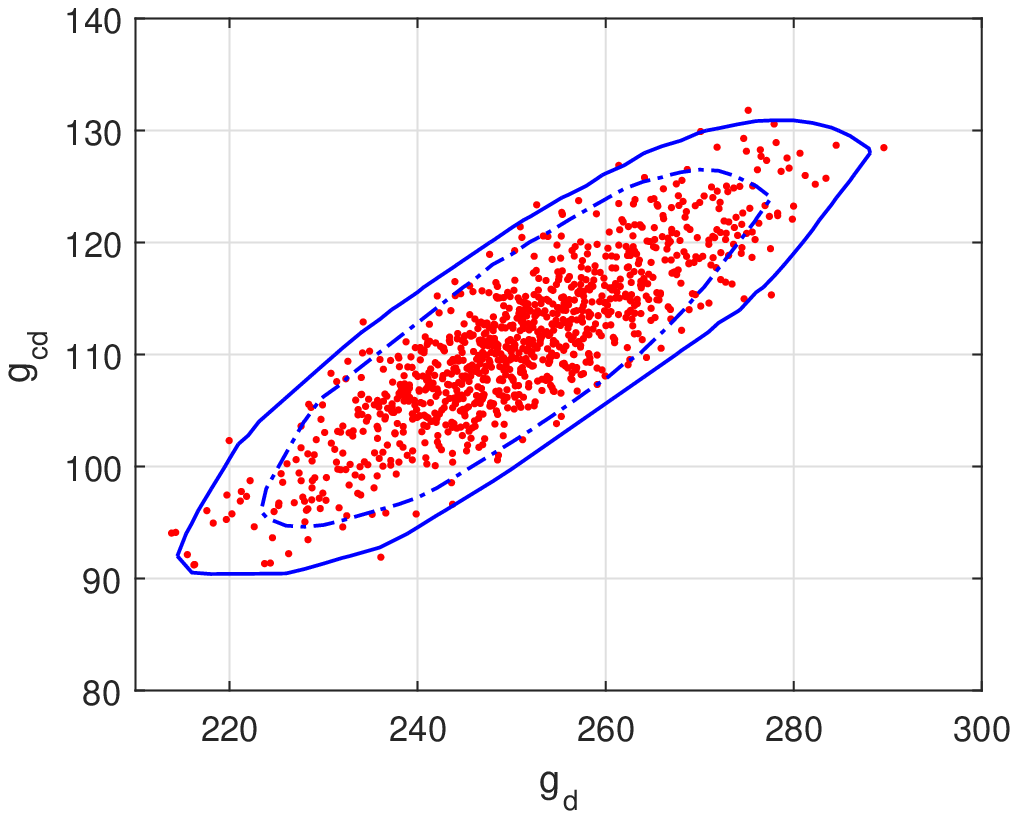}}
            \end{minipage}}
                \subfigure[SVC under under gaussian.]{
          \begin{minipage}[b]{0.3\linewidth}
            \centering
            \resizebox{\linewidth}{!}{%
              \includegraphics[height=0.5in]%
              {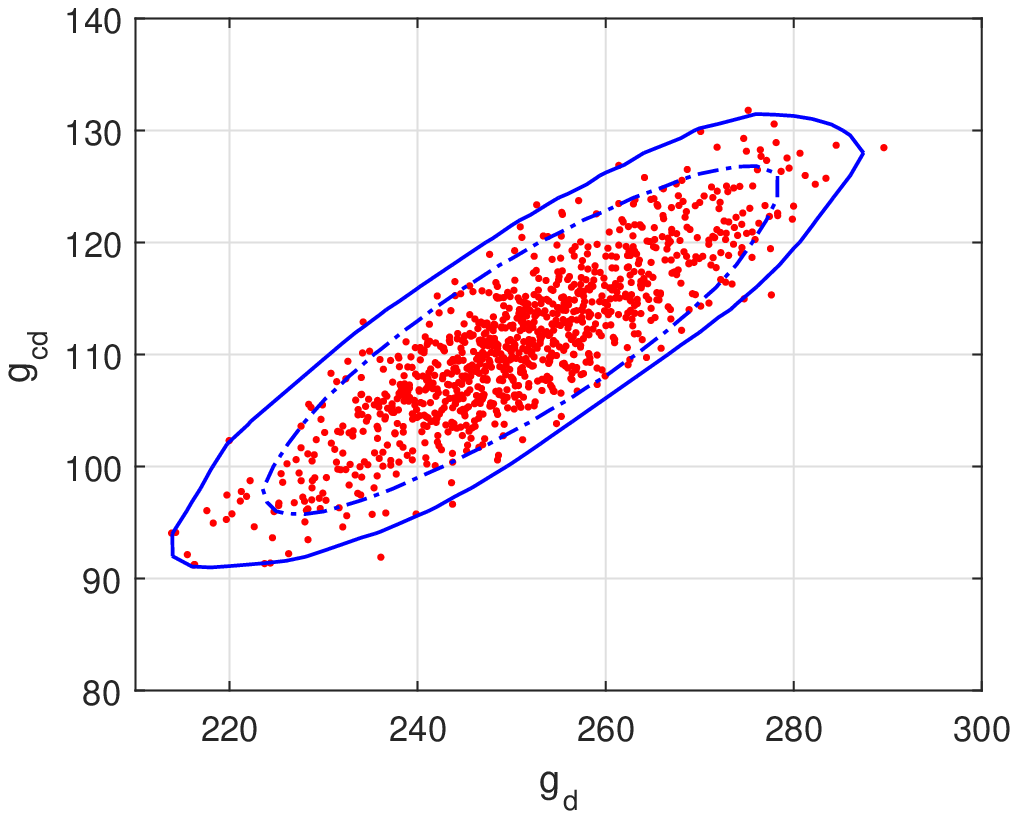}}
            \end{minipage}}
             \caption{Learning results of the uncertainty sets, where the solid line and the dotted line correspond to $\epsilon=0.01$ and $\epsilon=0.05$ respectively}
\end{figure*}

\subsection{Quantile-SVC Approach}

The Quantile-SVC approach adopts the hard margin to seek the smallest sphere to enclose the data in high-dimensional features space, which can be formulated as the following optimization problem
\begin{eqnarray}
\min\limits_{R,\bm\rho }&&   R^{2}   \\
 \textrm{s.t.} &&  \| \phi(\bm\xi_{i})-\bm\rho \|^{2}  \leq R^{2}, i=1,\cdots,N. \nonumber
\end{eqnarray}
Using the similar method in Section IV, the dual problem is computed as
\begin{eqnarray}
\min\limits_{\bm\lambda} && \sum_{i=1}^{N}\sum_{j=1}^{N}\lambda_{i}\lambda_{j}K(\bm\xi_{i},\bm\xi_{j})-\sum_{i=1}^{N}\lambda_{i}K(\bm\xi_{i},\bm\xi_{i})   \\
\textrm{s.t.}  &&  \sum_{i=1}^{N}\lambda_{i}=1,  \lambda_{i}\geq 0.  \nonumber
\end{eqnarray}
According to the KKT conditions and complementary slackness, when the $i$-th sample in the interior of the sphere $\|\phi(\bm\xi_{i})-\bm\rho \|^{2}<R^{2}$, we must have $\lambda_{i}=0$. On the contrary, when the $i$-th sample located on the boundary of the sphere $\|\phi(\bm\xi_{i})-\bm\rho \|^{2}=R^{2}$, there must exist $\lambda_{i}>0$. Note that only the sample $\bm\xi_{i}$ with positive $\lambda_{i}$ can contribute the construction of the center. We refer to this samples as support vectors and the index set can be expressed as
\begin{eqnarray}
\mathcal{S}_{v}=\{i\mid \lambda_{i}>0,\forall i \}.
 \end{eqnarray}
Besides, the center of sphere is also computed as a linear combination of the mapping of all support vectors, i.e.,  $\bm\rho=\sum_{i\in \mathcal{S}_{v}}\lambda_{i}\bm\phi(\bm\xi_{i})$. Then, we should calibrate the size of the sphere so that it can cover the uncertain CSI samples in high-dimensional features space with confidence $1-\epsilon$. For using the quantile estimation method, we define
\begin{eqnarray}
t_{svc}(\bm\xi_{l})=\|\phi(\bm\xi_{l})-\bm\rho\|
\end{eqnarray}
to map the channel sample into real number. Then, the size parameter $R^{2}$ will be chosen as $n_{\epsilon}=\lceil(1-\epsilon)N\rceil$-th value of the ranked observations $t_{svc}^{(1)}\leq\cdots\leq t_{scv}^{(N)}$ in ascending order, i.e., $R^{2}=t_{svc}^{(n_{\epsilon})}(\bm\xi_{l_{\epsilon}})$. After that, the feasible set of uncertain CSI can be defined as the sphere with radius $R^{2}$, i.e.
\begin{eqnarray}
\mathcal{S}_{\epsilon}(\mathcal{N})=\{\mathbf{g}\mid\phi(\mathbf{g})-\bm\rho\leq R^{2} \}.
\end{eqnarray}
Then, using the similar process as shown in Section IV, the robust counterpart of chance constraint based on Quantile-SVC approach can be computed as the following linear constraint
\begin{eqnarray}
\left\{\begin{array}{l}
         \sum_{i\in \mathcal{S}_{v}}(\bm\omega_{i}-\bm\varphi_{i})^{T}\mathbf{Q}\bm\xi_{i}-\varrho\kappa\geq \gamma_{min}^{d}, \\
         \sum_{i\in \mathcal{S}_{v}}(\bm\omega_{i}-\bm\varphi_{i})^{T}\mathbf{Q}-\mathbf{p}=\mathbf{0}, \\
         \bm\omega_{i}+\bm\varphi_{i}-\lambda_{i}\cdot\kappa\cdot\mathbf{1}=\mathbf{0},\forall i\in\mathcal{S}_{v}, \\
         \kappa\geq 0,\bm\varphi_{i},\bm\omega_{i}  \in \mathbb{R}^{2}_{+},
       \end{array}
\right.
\end{eqnarray}
where $\varrho=\sum_{i\in\mathcal{S}_{v}}\lambda_{i} \|\mathbf{Q}(\bm\xi_{l_{\epsilon}}-\bm\xi_{i})\|_{1}$. Similar to the robust counterpart in Eq. (47), the number of edges of the obtained polytope in (57) also depends on the number of support vectors in $\mathcal{S}_{v}$. However, it can be known from the computation of the dual problem in (57) that the number of support vectors has nothing to do with the outage probability requirement $\epsilon$. That means that the number of edges of the obtained polytope in (57) is the same for all possible values of $\epsilon$. From the quantile estimation in (55), we knows that the change of $\epsilon$ leads to the case that different CSI sample $\bm\xi_{l_{\epsilon}}$ will be selected to compute the size of the sphere in high-dimensional features space. As follows, the parameter $\varrho=\sum_{i\in\mathcal{S}_{v}}\lambda_{i} \|\mathbf{Q}(\bm\xi_{l_{\epsilon}}-\bm\xi_{i})\|_{1}$, which can be considered the size parameter of the polytope, will change so that the obtained polytope can cover the uncertain CSI with confidence $1-\epsilon$.

\subsection{Experiment Results}

In the first experiment, we model the uncertain CSI using the proposed learning approaches with different outage probabilities, and the graphical results are shown in Fig. 4. We observe from Figs. 4(a) and (b) that for covering the uncertain CSI with the specific confidence level, the sysmmetrical-geometry-based uncertainty set will result in large superfluous coverage. This potential leads to over-conservative power allocation solutions. However, it is shown in Figs. 4(c)-(f) that both Quantile-SVC and SVC yield the convex and asymmetric uncertainty sets which can compactly cover the uncertain CSI without too much superfluous coverage. This verifies the effectiveness of the ML technique in clustering the uncertain CSI samples. From these figures, we also observe that when the outage probability $\epsilon$ is set as 0.01, regardless of the distribution of the uncertain CSI, the polytope shapes learned by Quantile-SVC and SVC are almost the same. However, with the increase of outage probability, the polytope learned by Quantile-SVC approach will shrink proportionally without changing its shape. This verifies our analysis about the robust counterpart in (57). For the SVC approach, when the outage probability increases, the size of the enclosing envelop decreases, while the number of edges of the uncertainty set increases. As a result, the rim of the polytope tends to be more and more smooth. Thus, more samples are considered as outliers residing outside the uncertainty set and the induced polytope becomes less conservative.

In Figs. 5 and 6, we evaluate the convergence of the bisection search-based power allocation algorithm. The figures show that the proposed algorithm converges within only at most 17 iterations. This verifies the low complexity of the proposed algorithm. We also observe from the figures that no matter under which robust optimization approach, the transmit power of CUE can reach the maximum value. This verifies the solution in \textbf{Lemma 1}. Besides that, Fig. 6 shows that the DUE transmit powers under Box, Ellipsoidal and Polytope approaches are much larger than the ones under SVC, Quantile-SVC and Non-Robust. That is because all of the sysmmetrical-geometry-based uncertainty sets will result in superfluous coverage. For satisfying the chance D2D QoS constraint, the DUE with the deployment of these approaches must consume more transmit power than others. Benefiting from the advanced uncertainty set learning method, the DUE with  SVC and Quantile-SVC can satisfy the QoS constraint with smaller transmit powers. Although the Non-Robust approach achieves the smallest transmit power, the chance D2D QoS requirement is not necessarily guaranteed.

%\begin{figure}[htbp]
%        \centering
%        \subfigure[\scriptsize {Media content flows rates.}]{
%          \begin{minipage}[b]{0.45\linewidth}
%            \centering
%            \resizebox{\linewidth}{!}{%
%              \includegraphics[height=1.0in]%
%              {CUE_power}}
%          \end{minipage}}
%        \subfigure[\scriptsize{Wireless links transmission rates.}]{
%          \begin{minipage}[b]{0.45\linewidth}
%            \centering
%            \resizebox{\linewidth}{!}{%
%              \includegraphics[height=1.0in]%
%              {DUE_power}}
%            \end{minipage}}
%             \caption{Media streaming decisions under $V=50$ and $\alpha=0.55$.}
% \end{figure}

\begin{figure}
\begin{center}
\includegraphics[width=3.0in,height=2.4in]{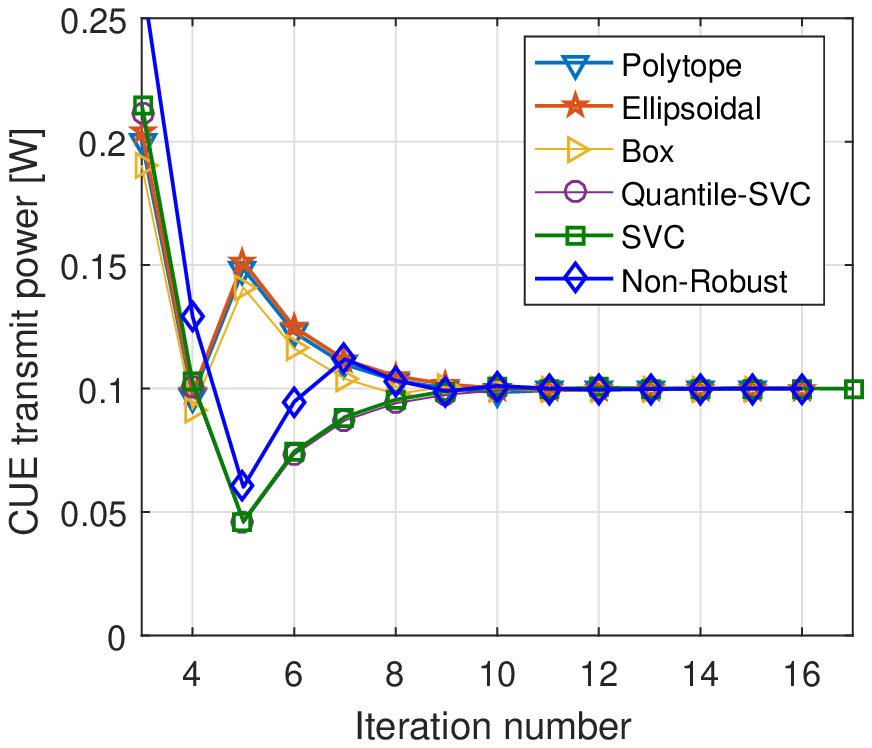}
\label{fig2}\caption{Convergence of the bisection search-based algorithm, assuming $P_{max}^{c}=P_{max}^{d}=20$dBm, QoS requirements $\gamma_{min}^{c}=5$, $\gamma_{min}^{d}=0.1$ and outage probability $\epsilon=0.05$.}
\end{center}
\end{figure}

\begin{figure}
\begin{center}
\includegraphics[width=3.0in,height=2.4in]{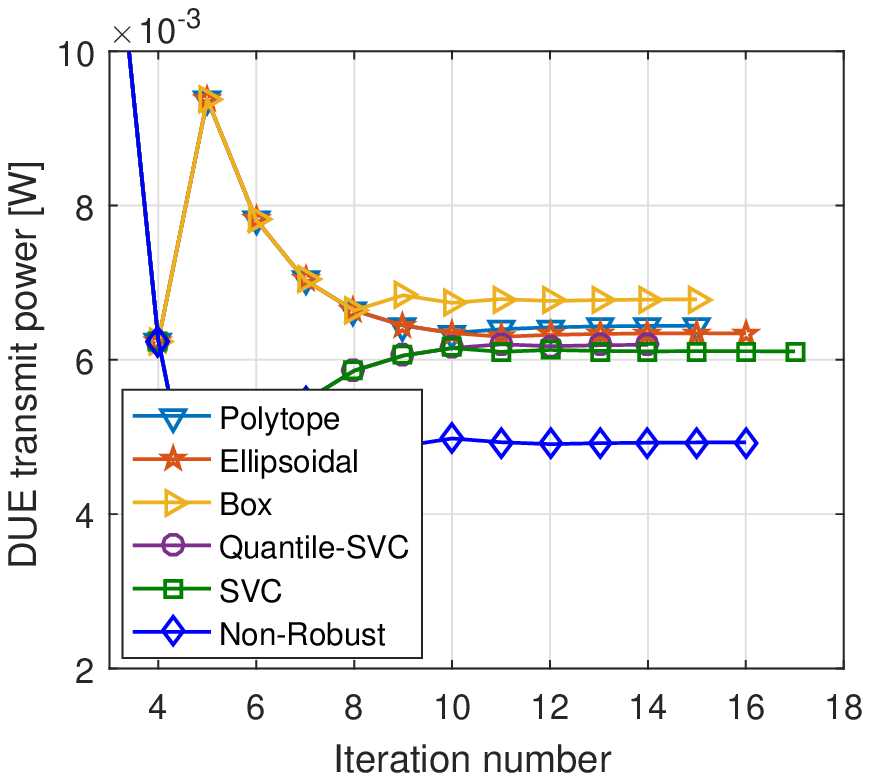}
\label{fig2}\caption{Convergence of the bisection search-based algorithm, assuming $P_{max}^{c}=P_{max}^{d}=20$dBm, QoS requirements $\gamma_{min}^{c}=5$, $\gamma_{min}^{d}=0.1$ and outage probability $\epsilon=0.05$.}
\end{center}
\end{figure}

\begin{figure}
\begin{center}
\includegraphics[width=3.0in,height=2.4in]{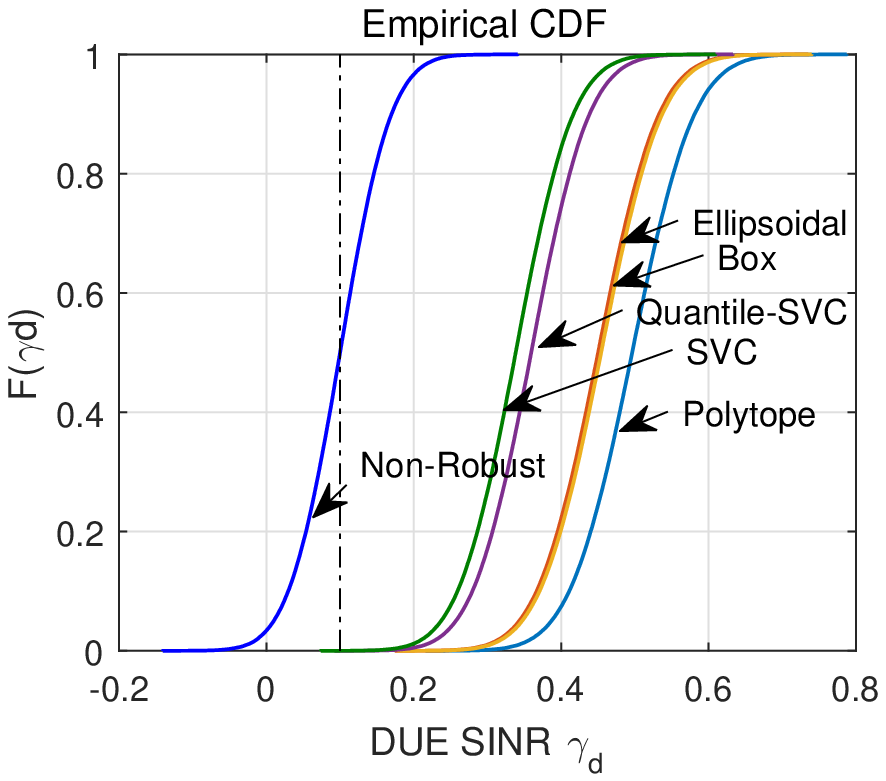}
\label{fig2}\caption{DUE SINR cumulative distributions under Gaussian uncertainties, assuming $P_{max}^{c}=P_{max}^{d}=20$dBm, QoS requirements $\gamma_{min}^{c}=5$, $\gamma_{min}^{d}=0.1$ and outage probability $\epsilon=0.05$.}
\end{center}
\end{figure}

\begin{figure}
\begin{center}
\includegraphics[width=3.0in,height=2.4in]{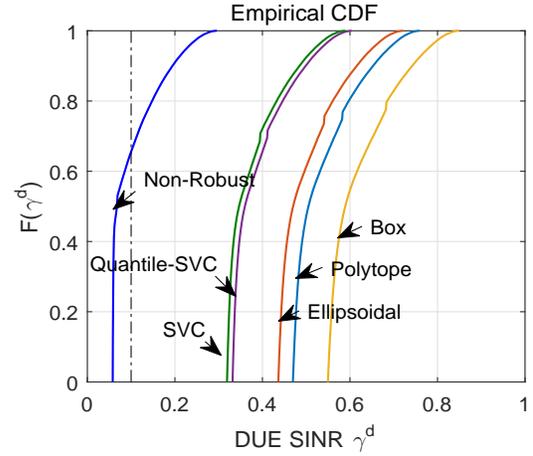}
\label{fig2}\caption{DUE SINR cumulative distributions under exponential uncertainties, assuming $P_{max}^{c}=P_{max}^{d}=20$dBm, QoS requirements $\gamma_{min}^{c}=5$, $\gamma_{min}^{d}=0.1$ and outage probability $\epsilon=0.05$.}
\end{center}
\end{figure}

Figs. 7-9 illustrate the DUE SINR cumulative distributions for different robust optimization approaches under the test set. It is not difficult to understand that for all of the curves, the point $(0.1,F(0.1))$  can be considered as the outage probability of D2D communication. Then, we observe from the figures that no matter what the distribution of the channel error is, the outage probability of Non-robust approach is almost larger than 0.5. Such terrible performance is very dangerous for many application scenarios. For example, the reliability requirement for D2D-enabled vehicular communications at the vehicle platooning use case can reach up to 99\% \cite{G.NaiKACCESS}. On the contrary, by substituting the chance constraint as the proposed robust counterparts, all of the robust optimization approaches achieve satisfactory outage performance. However, we should note that the achieved SINR of Box, Polytope and Ellipsoidal is almost larger than 0.3 for the Gaussian uncertainties and 0.4 for the exponential uncertainties. This performance is too conservative, and will lead to the waste of wireless resources. Thus, by using the ML method on learning the uncertainty set, the conservatism can be overcame very well. We also observe from the figures that the achieved SINR under Quantile-SVC is larger than the one under SVC. This shows that Quantile-SVC is more conservative than SVC. This phenomenon can be explained in two aspects. First, at different outage probabilities, the polytope specified by the robust counterpart of Quantile-SVC can only shrink proportionally. However, the envelope of the polytope specified by SVC tends to be more and more smooth when the outage probability increases. Thus, SVC shows higher flexibility than Quantile-SVC, so it has better ability for overcoming the conservatism. Second, we note that the size of uncertainty set in Quantile-SVC is estimated by using the quantile estimation method in high-dimensional features space. Recalled the definition of $\phi(\bm\xi)$, the uncertain CSI from two-dimensional space is mapped into to three-dimensional space by the nonlinear mapping. Therefore, the order about the radius of sphere in high-dimensional space may not be applicable in the two-dimensional CSI space. This lead to the result that the polytope in two-dimensional CSI space cannot cover the uncertain CSI samples with confidence level as $1-\epsilon$, so that it is more conservative than SVC.

\begin{figure}
\begin{center}
\includegraphics[width=3.0in,height=2.4in]{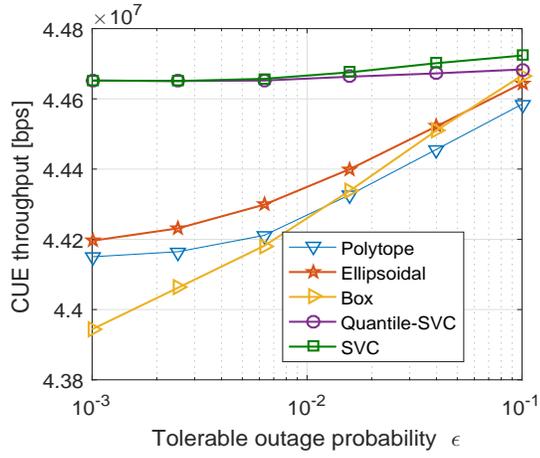}
\label{fig2}\caption{CUE throughput versus tolerable outage probability, assuming $P_{max}^{c}=P_{max}^{d}=20$dBm and QoS requirements $\gamma_{min}^{c}=5$, $\gamma_{min}^{d}=0.1$.}
\end{center}
\end{figure}

\begin{figure}
\begin{center}
\includegraphics[width=3.0in,height=2.4in]{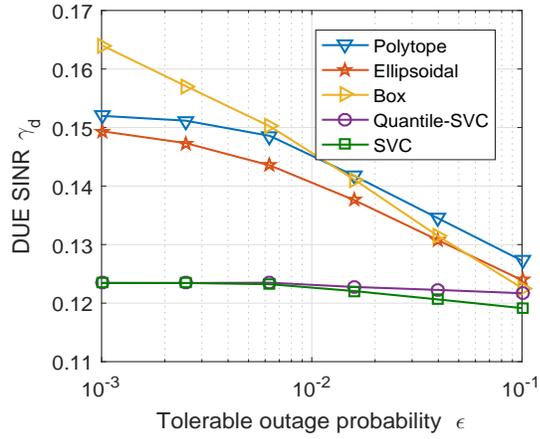}
\label{fig2}\caption{DUE SINR versus tolerable outage probability, assuming $P_{max}^{c}=P_{max}^{d}=20$dBm and QoS requirements $\gamma_{min}^{c}=5$, $\gamma_{min}^{d}=0.1$.}
\end{center}
\end{figure}

In Figs. 9 and 10, we illustrate the CUE throughput and the DUE SINR versus the tolerable outage probability. Because all of the robust optimization approaches show the similar performance under different uncertainty distribution, thus this and the following experiment only exhibit the results under truncated exponential distribution. In addition, both the CUE throughput and the DUE SINR under Non-Robust approach are much worse than the ones under other optimization approaches and it is difficult to exhibit them in the same figure. Thus, the performance of Non-Robust approach will not be shown any more in this and the following experiments. In Figs. 9 and 10, we observe that when the tolerable outage probability is small, the symmetrical-geometry-based robust optimization approaches obtain the smaller CUE throughput than others. This is because these approaches are very conservative. They greedily consume a lot of power to guarantee the chance constraint of D2D. As a result, the obtained CUE throughput is smaller than others. On the contrary, the ML-based robust optimization approaches overcome the conservatism very well so that they achieve higher CUE throughput whilst ensuring the chance QoS constraint of D2D. The figures also show that with the increase of tolerable outage probability, the DUE SINR under all of the robust optimization approaches decreases while the CUE throughput increases. This is because the size of uncertainty set  decreases with the increase of tolerable outage probability, so the DUE no longer needs to consume much power for guaranteeing the chance QoS requirement. As a result, the obtained CUE throughput increases.

\begin{figure}
\begin{center}
\includegraphics[width=3.0in,height=2.4in]{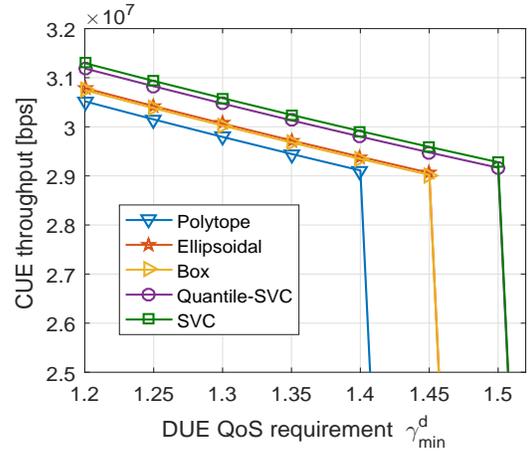}
\label{fig2}\caption{CUE throughput versus DUE QoS requirement, assuming $P_{max}^{c}=P_{max}^{d}=20$dBm, CUE QoS requirement $\gamma_{min}^{c}=5$ and outage probability $\epsilon=0.05$.}
\end{center}
\end{figure}

\begin{figure}
\begin{center}
\includegraphics[width=3.0in,height=2.4in]{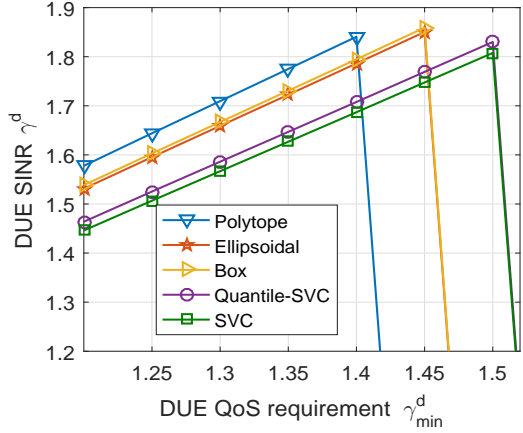}
\label{fig2}\caption{DUE SINR versus DUE QoS requirement, assuming $P_{max}^{c}=P_{max}^{d}=20$dBm, CUE QoS requirement $\gamma_{min}^{c}=5$ and outage probability $\epsilon=0.05$.}
\end{center}
\end{figure}

Figs. 11 and 12 illustrate the CUE throughput and the DUE SINR versus the DUE QoS requirement $\gamma_{min}^{d}$. We observe from these figures that, with the increase of DUE QoS requirement, the achieved DUE SINR increases while the CUE throughput decreases. This is because the DUE needs to consume more transmit power to support the increased DUE SINR. As a result, its interference to the CUE increases so that the CUE throughput decreases. The figures also show that at under any DUE QoS requirement, the ML-based robust optimization approach can overcome the conservatism of the symmetrical-geometry-based approaches so that they achieve the larger CUE throughput. Moreover, due the effectiveness of SVC on the representations of sample distribution, it always shows better performance on the DUE SINR and achieves larger CUE throughput than Quantile-SVC. We also observe from the figures that when the DUE QoS requirement is larger than 1.4, the proposed robust optimization approaches will collapse one after another. This is because the total power resources of the network are limited, so they cannot support such a large D2D QoS requirement. Because the symmetrical-geometry-based approaches consume too much power for protecting the chance constraint, so that they collapse earlier than the others. On the contrary, both SVC and Quantile-SVC can support a larger DUE QoS requirement. This illustrates the effectiveness of the ML-based robust optimization approaches.

\begin{figure}
\begin{center}
\includegraphics[width=3.0in,height=2.4in]{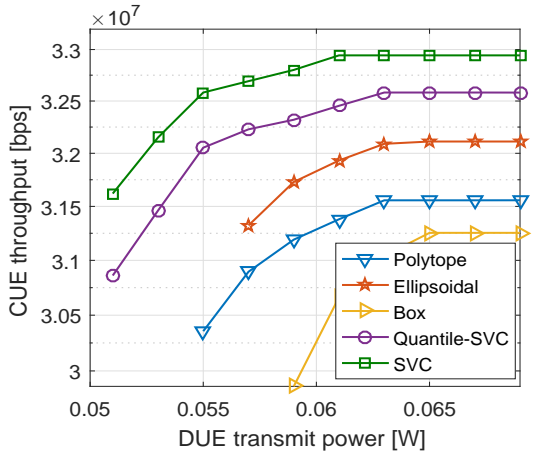}
\label{fig2}\caption{CUE throughput versus DUE transmit power, assuming $P_{max}^{c}=20$dBm, CUE QoS requirement $\gamma_{min}^{c}=5$, DUE QoS requirement $\gamma_{min}^{c}=1$ and outage probability $\epsilon=0.05$.}
\end{center}
\end{figure}

\begin{figure}
\begin{center}
\includegraphics[width=3.0in,height=2.4in]{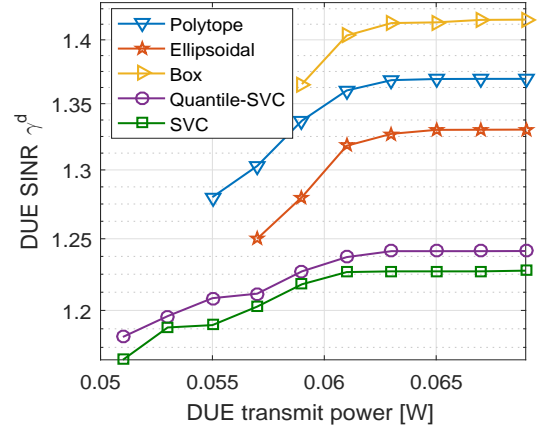}
\label{fig2}\caption{DUE SINR versus DUE transmit power, assuming $P_{max}^{c}=20$dBm, CUE QoS requirement $\gamma_{min}^{c}=5$, DUE QoS requirement $\gamma_{min}^{c}=1$ and outage probability $\epsilon=0.05$.}
\end{center}
\end{figure}

In Figs. 13 and 14, we illustrate the CUE throughput and the DUE SINR versus the DUE transmit power. In these figures, we observe that when the DUE transmit power is small, all of the robust optimization approaches cannot find the feasible solution for the power allocation problem. With the increase of DUE transmit power, the ML-based robust optimization algorithms revive first, followed by the symmetrical-geometry-based approaches. This is because the ML-based robust optimization approaches can overcome the conservatism of the symmetrical-geometry-based approaches, so they need the smaller transmit power than others for finding the feasible solution to satisfy the D2D QoS requirement. Moreover, we also observe from the figures that although the Quantile-SVC approach can revive together with SVC, its CUE throughput performance is worsen than SVC. Figs. 15 and 16 plot the CUE throughput and DUE SINR versus the channel estimation error coefficient. As the statements of channel model in Section II, with the decrease of channel estimation error coefficient, the channel uncertainty gets more and more larger. Thus, the DUE need to allocate more transmit power to get the larger SINR for satisfying the chance QoS constraint. As a result, as shown in Fig. 16, the achieved DUE SINR of all approaches increases. Undoubtedly, the increased DUE transmit power will produce serious interference to CUE. Thus, we observe from Fig. 15 that the CUE throughput decreases with the decrease of channel estimation error coefficient. Besides, Figs. 15 and 16 illustrate the similar phenomenon as Figs. 11 and 12. With the increase of channel uncertainty, the robust optimization approaches will collapse one after another. However, benefiting from the effectiveness of the ML-based robust optimization on representation  of the uncertainty distribution, both SVC and Quantile-SVC approaches can deal with larger channel uncertainty than the symmetrical-geometry-based robust optimization approaches. This verifies the effectiveness of the proposed ML-based robust optimization approaches again from another perspective.

\begin{figure}
\begin{center}
\includegraphics[width=3.0in,height=2.4in]{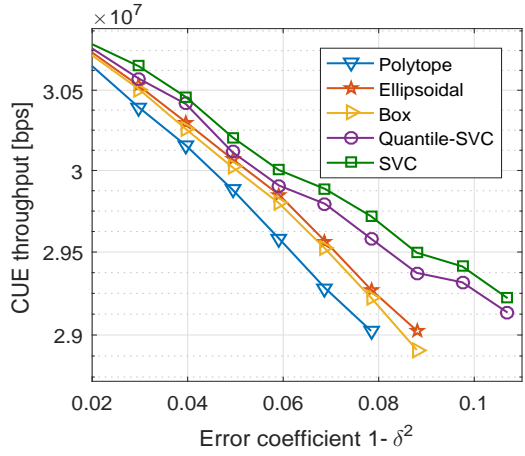}
\label{fig2}\caption{CUE throughput versus channel estimation error coefficient, assuming $P_{max}^{c}=P_{max}^{d}=20$dBm, CUE QoS requirement $\gamma_{min}^{c}=5$, DUE QoS requirement $\gamma_{min}^{c}=1.5$ and outage probability $\epsilon=0.05$.}
\end{center}
\end{figure}

\begin{figure}
\begin{center}
\includegraphics[width=3.0in,height=2.4in]{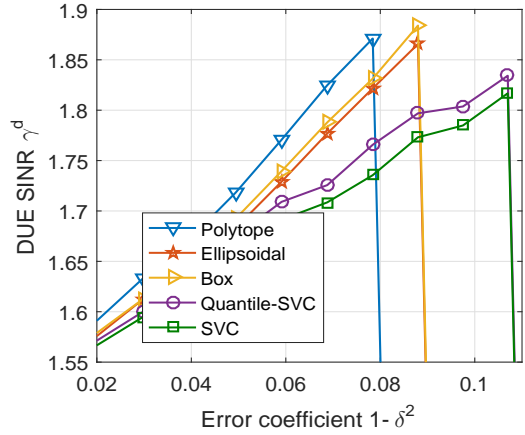}
\label{fig2}\caption{DUE SINR versus channel estimation error coefficient, assuming $P_{max}^{c}=P_{max}^{d}=20$dBm, CUE QoS requirement $\gamma_{min}^{c}=5$, DUE QoS requirement $\gamma_{min}^{c}=1.5$ and outage probability $\epsilon=0.05$.}
\end{center}
\end{figure}

\section{Conclusions}

In this paper, we studied the resource allocation in D2D underlaying cellular network. The problem was formulated as maximizing the CUE throughput under the channel uncertainties whilst guaranteeing a minimum SINR requirement constraint for D2D. We proposed a robust resource allocation framework for solving the highly intractable chance constraint about D2D service requirement. Then, we modeled the uncertain CSI into polytope, ellipsoidal and box and derived the robust counterparts of the chance constraint under these uncertainty sets. To overcome their conservatism, we developed a support vector clustering (SVC)-based approach to model uncertain CSI as a compact convex uncertainty set. Finally,  we developed a bisection search-based power allocation algorithm for solving the resource allocation in D2D underlaying cellular network with different robust counterparts.

\footnotesize
\small
\bibliographystyle{IEEEtran}
%\bibliography{references}

\end{document}